\newif\ifDRAFTMODE
\ifDRAFTMODE
	\documentclass[journal, a4paper, draftclsnofoot, onecolumn]{IEEEtran}
\else
	\documentclass[journal, a4paper, final, twocolumn]{IEEEtran}
\fi

\usepackage{amsmath, epsfig, times, amsthm, amsfonts, amssymb, comment, subfigure, multirow, algorithm, algorithmic}

\usepackage{soul}

\usepackage{color}

\newtheorem{prop}{Proposition}
\newtheorem{corol}{Corollary}

\begin{document}
\title{Band Codes for Energy-Efficient Network Coding with Application to P2P Mobile Streaming}

\author{
	Attilio~Fiandrotti,~\IEEEmembership{Member,~IEEE,}
	Valerio~Bioglio,~\IEEEmembership{Member,~IEEE,}
	Marco~Grangetto,~\IEEEmembership{Senior Member,~IEEE,}
	Rossano~Gaeta,
	and~Enrico~Magli,~\IEEEmembership{Senior Member,~IEEE}
  \thanks{
  Copyright (c) 2013 IEEE. Personal use of this material is permitted. However, permission to use this material for any other purposes must be obtained from the IEEE by sending a request to pubs-permissions@ieee.org.
  
  This publication is based partly on work performed within Project COAST-ICT-248036 which is funded by the European Union, partly on work performed within project AMALFI which is funded by Università di Torino and Compagnia di San Paolo, partly on work performed within project ARACNE, a PRIN funded by the Italian Ministry of Education and Research.

  A.Fiandrotti, V.Bioglio, and E. Magli are with the Department of Electronics and Telecommunications, Politecnico di Torino, 10129, Torino, Italy (e-mail: attilio.fiandrotti@polito.it; valerio.bioglio@polito.it; enrico.magli@polito.it).

 M.Grangetto and R.Gaeta are with the Department of Computer Science, Universit\`a di Torino, 10149 Torino, Italy  (e-mail: marco.grangetto@di.unito.it; rossano.gaeta@di.unito.it).
  }
}

\maketitle

\begin{abstract}

A key problem in network coding (NC) lies in the complexity and energy consumption associated
with the packet decoding processes, which hinder its application in mobile environments. Controlling and hence limiting such factors has always been an important but elusive research goal, since the packet degree distribution, which is the main factor driving the complexity, is altered in a non-deterministic way by the random recombinations at the network nodes. In this paper we tackle this problem with a new approach and propose Band Codes (BC), a novel class of network codes specifically designed to preserve the packet degree distribution during packet encoding, recombination and decoding. BC are random codes over GF(2) that exhibit low decoding complexity, feature limited and controlled degree distribution by construction, and hence allow to effectively apply NC even in energy-constrained scenarios. In particular, in this paper we motivate and describe our new design and provide a thorough analysis of its performance.We provide numerical simulations of the BC performance in order to validate the analysis and assess the overhead of BC with respect to a conventional random NC scheme. Moreover, experiment in a real-world application, namely peer-to-peer mobile media streaming using a random-push protocol, show that BC reduce the decoding complexity by a factor of two with negligible increase of the encoding overhead, paving the way for the application of NC to power-constrained devices.

\end{abstract}


% \fi
\IEEEpeerreviewmaketitle

\section{Introduction}
\label{sec:introduction}

Network Coding (NC)~\cite{fragouli2006network} has attracted a lot of interest recently due to its potential to maximize the network throughput in multicast communications. 
In a typical NC scenario~\cite{ahlswede2000network}, a source node wants to share a message, also known as \emph{generation}, with multiple nodes in the network. 
The source divides the message in input symbols of a given size and transmits linear combinations of these symbols (termed as \emph{encoded packets}) to the nodes of the network.
Each network node receives encoded packets and transmits linear combinations thereof to the other network nodes.
Once a node has collected enough linearly independent encoded packets, it solves a system of linear equations and recovers the message.
The recombinations at the network nodes are the key to better network throughput, as they increase the probability that a packet is \emph{innovative} at its recipient, i.e. that it is useful to recover the message.

NC brings several benefits to cooperative media streaming.
First, the network nodes start transmitting packets before they have recovered the message, which is a major edge in delay-sensitive applications such as media streaming.
Media streaming is in fact a challenging application because reduced delays and constant throughput are required to achieve smooth playback.
Second, the network nodes can be arranged as totally random overlays, drastically reducing the need for coordination among the network nodes in peer-to-peer (P2P) communications.
Wang \emph{et al.} showed the benefits of NC for collaborative P2P video distribution with their streaming protocol $R^2$~\cite{wang2007r2, wang2007network}.
In their protocol, the packet scheduler operates according to a random-push packet mechanism and the peers are organized as a random overlay.
Their experiments showed that NC enables better video quality thanks to improved network throughput and reduced buffering times.
%The protocol was evaluated on a testbed composed of an array of high-end workstations that simulated a live video streaming session to a large population of users.
%Their experiments on a controlled testbed showed that NC allow sustained throughput and reduced buffering delays, enabling high quality streaming of live video contents.
%Our experiments~\cite{toldo2010resilient} with a push-based video streaming architecture based on multiple trees showed reduced buffering times together with improved resilience to ungraceful peers departures.
%In our previous work~\cite{fiandrotti2011complexity}, we experimentally verified that a random-push P2P video streaming protocol of our design could reduce the required buffering times with respect to the NextShare protocol~\cite{nextshare}.
%This witnesses the gains that NC offers to improve the performance of multimedia communications applications.

The increasing popularity of devices such as mobile phones and tablets is fostering the demand for energy-efficient media delivery architectures. %, the computational complexity of NC is an issue given the battery-powered nature of mobile devices.
In the future, hybrid media distribution architectures based on NC are envisioned, where terminals of different types, for example mobile phones and PCs, cooperate to distribute video contents.
NC is however an energy-demanding application, especially when the coding operations are performed over high order fields such as $GF(2^8)$, hence the need for energy-efficient NC schemes.
%The interest towards NC has spurred a lot of research towards energy-efficient NC schemes that could be practical even for mobile devices.
%Rateless codes have been proposed as a solution to reduce the computational complexity of NC.
A straightforward way to reduce the NC computational complexity independently from the field size is to reduce the size of the generation, i.e. resorting to messages composed of fewer symbols.
However, as to smaller messages correspond lower efficiency of the code, generation of higher size are usually desirable.
Moreover, in multimedia applications the size of the message cannot be arbitrarily reduced but is rather bounded by the characteristics of the underlying media stream.
Existing research in low-complexity erasure correction codes hints the way towards energy-efficient NC.
Rateless codes~\cite{Mackay05fountaincodes} such as LT~\cite{LT} and Raptor~\cite{raptor} codes are low-complexity erasure correction codes defined over $GF(2)$.
The asymptotic performance of rateless codes is not far from that of coding schemes defined over larger fields, but their decoding complexity is much lower as their require a controlled number of simpler XOR operations.
In a typical source-receiver scenario, the source draws the number of symbols to encode in each packet, also known as packet \emph{degree}, according to a purposely designed distribution such as the RSD distribution used in LT codes.
The packet degree distribution controls the trade-off between encoding efficiency, i.e. the number of coded packets that a node needs to receive to reconstruct the original message, and decoding complexity, i.e. the number of XOR operations and hence the energy required to recover the message.
%Lucani et al.~\cite{lucani2009random}, \cite{lucani2010systematic} investigated NC with rateless codes and showed that the rateless codes performance is not far from that of NC schemes defined over larger fields such as $GF(2^8)$, albeit using fewer and simpler XOR operations.
%Similarly, our previous work~\cite{fiandrotti2011complexity} showed that NC over $GF(2)$ enables sustained throughput and low delay at the cost of a rather small bandwidth overhead.
However, in a NC scenario the recombinations at the network nodes alter the packet degree distribution selected at the source.
Therefore, the decoding complexity increases at each recombination as described in detail in Section~\ref{sec:rnc_over_gf2}. %and that is the problems that we tackle in this work.
Several approaches to this problem have been proposed as described in detail in Section~\ref{sec:related}, however none of them completely satisfies the requirements for random-push P2P video distribution over random overlays.

%In this paper we show how the joint design of the encoding at the source and recombination process at the receivers enables energy-efficient NC, and we show its application to mobile communications.
\subsection*{Our contributions}
Our contributions towards energy-efficient, low-complexity, NC are in the following.
\begin{itemize}

\item  We show analytically that random recombinations at the nodes of a random graph lead to an uncontrolled increase of the decoding complexity.  %with the number of recombinations.
% how the recombinations at the nodes alter the degree distribution of the network packets and bloat the decoding complexity of the nodes.
%While intuitive explanations have been proposed before, in this work we provide a sound demonstration of the problem.

\item We provide a thorough description of \emph{Band Codes} (BC), a novel class of network codes, preliminary described in~\cite{icme2012BC}.
BCs achieve controlled decoding complexity thanks to the joint design of the encoding,  decoding, and recombination processes.
The most important feature of BCs  is that the proposed
recombination strategy does not change the packet degree distribution after an arbitrary number 
of recombinations.
%A mathematical proof of this property is given.
As a consequence,
BC can be generated at a given source and recombined with a simple mechanism at intermediate
nodes without impacting on the degree distribution, that in turns determines both the decoding computational cost and the encoding efficiency. Clearly, the actual coding efficiency in a P2P network coding scenario depends on the particular overlay topology. 
In this paper we show that in a random mesh topology that contains cycles the BC packet recombination algorithm yields good coding efficiency.

\item  An analytical model of BC decoding complexity is derived allowing to match the
decoding computational cost to the the capacity of the device at hand. 
%While BC are designed for NC over random networks with cycles, they preserve the packet degree distribution even with arbitrary network topologies.
%Also, the decoding process is exploited as a preliminary recombination stage~\cite{fiandrotti2011complexity}, which reduces the complexity of the packet recombination stage without impacting the encoding efficiency.

\item We exploit the BC features in a P2P video streaming application using the random-push protocol  described in~\cite{fiandrotti2011complexity}.
%The distinguishing feature of the protocol is the ability of each node of the network to independently control its own computational complexity exploiting the properties of BC, enabling the coexistence of peers with different computational capabilities within the same P2P session.
This has allowed us to perform an extensive experimentation to asses the encoding efficiency, the decoding complexity, and the energy consumption of BC enabled devices in a practical use case.
We have worked out experiments in both controlled scenarios including a set of desktop computers and mobile phones in our laboratory and on a planetary scale set-up using the Planetlab network.

%properties of BC in terms of encoding efficiency, decoding complexity and resulting energy consumption and on the Internet to evaluate the video quality that BC enable in a real network.
\end{itemize}

The remainder of this paper is organized as follows.
In Section~\ref{sec:related} we overview the existing literature on the topic of low-complexity NC with a special eye to mobile applications.
In Section~\ref{sec:rnc_over_gf2} we overview NC in low-order Galois Fields and we analytically describe how random recombinations at the nodes alter the packet degree distribution and increase the decoding complexity.
In Section~\ref{sec:proposed_architecture} we present BC, while in Section~\ref{sec:protocol} we describe their application to P2P video streaming.
Finally, Section~\ref{sec:rnc_experiments} provides an experimental evaluation of BC in terms of computational complexity, encoding efficiency and  video quality.
In Section~\ref{sec:conclusions} we draw the conclusions and outline possible future developments of this work.

\section{Related Work}
\label{sec:related}

In this section, we overview the existing literature on energy-efficient and low-complexity NC for mobile communications.
In its best known form, NC is defined over finite fields such as $GF(2^8)$, as the size of the field guarantees that the packets received by the nodes are innovative with high probability.
Due to the computational burden of directly computing multiplications on $GF(2^q)$, a common approach is to replace multiplications with additions by means of Look Up Tables tables (LUTs).
In~\cite{toldo2010}, for example, LUTs of the size of 256 bytes were used to solve multiplications over $GF(2^8)$.
However, the latency penalty in accessing LUTs stored in main memory may throttle the decoder throughput, especially on mobile devices where the constraints on power consumption impose to design processors with small caches.
Vingelmann \emph{et al.}~\cite{vingelmann2010} measured the decoder throughput achievable on an iPhone for NC over $GF(2^8)$ and $GF(2)$.
Their experiments showed that the best decoder throughput that could be achieved over $GF(2^8)$ was several times lower than over $GF(2)$.
%Previous investigations of ours produced similar outcomes and the evidence that the memory access latency in embedded devices are quite high due to the small size of the data caches.
Shojania \emph{et al.}~\cite{shojania2009random} considered NC-based video streaming with iPhone devices and their experiments showed that the coding operations accounted for more than 50\% of the processor cycles.
Heide at al.~\cite{heide2008cautious} also performed video streaming experiments with high-end mobile phones and went further measuring the impact of NC on the lifetime of the device.
Their experiments demonstrated not only that in some configurations the processor could not process the received packets fast enough resulting in a bottleneck with respect to the bandwidth available on the channel, but also found out that packet decoding reduced the operational life of the devices.
Angelopoulos \emph{et al.}~\cite{Angelopoulos:2011:EHI:2039912.2039928} showed that energy efficient NC is feasible also on mobile devices, but only at the price of using ad-hoc designed hardware. % to the device.
Concluding, the existing literature shows that NC over $GF(2^8)$ results in low decoder throughput and high computational loads on mobile devices, prompting the research for computationally lighter solutions.

A first step towards low complexity NC is to resort to simpler coding operations over $GF(2)$, also known as binary NC.
The main advantage of binary NC is that multiplications and additions are resolved with simple XORs, where a XOR is executed in one processor clock cycle avoiding the latency penalties associated with the use of lookup tables.
%Lucani ~\emph{et al.}~\cite{lucani2010systematic} showed that binary NC can provide a performance similar to that which could be achieved using larger fields in the context of systematic linear coding.
Moreover, the source can exploit rateless codes to create encoded packets to ensure low encoding and decoding complexity. 
Lucani et al.~\cite{lucani2009random, lucani2010systematic} investigated NC over $GF(2)$ and showed that the performance is not far from that of NC schemes defined over larger fields such as $GF(2^8)$, albeit using fewer and simpler XOR operations.
Katti~\emph{et al.}~\cite{katti2008xors} proposed the use of binary NC for packet routing in wireless networks, showing the sustained throughput together with low computational load that could be achieved.
However, in order to achieve low complexity NC, it is also important to control the number of XOR operations required to recover the message.
Although the appropriate selection of the packet degree at the source is enough to control the decoding complexity for erasure correction purposes, in a NC context the random recombinations at the network nodes alter the degree distribution as explained in the following Section.
So far, several solutions have been proposed to control the packet degree distribution in binary NC. %, here are the most significant.
%In~\cite{fiandrotti2011complexity} we proposed to exploit the decoder as a preliminary recombination stage to reduce the computational complexity of the recombination process.
Puducheri \emph{et al.}~\cite{puducheri} considered NC with LT codes and proposed a scheme such that the recombinations at the nodes do not alter the original degree distribution of LT codes.
However, this approach applies only to a network topology where one node recombines the packets received from two sources and relays to a single sink, and hence cannot be extended to arbitrary network topologies.
Thomos and Frossard~\cite{thomos2011degree} explored NC with Raptor codes and proposed an optimization framework to control the degree distribution at the source as a function of the network topology.
However, the requirement that the topology of the network is known at the source makes this scheme impractical where the network topology may be unknown at the source, as in mesh P2P communications, or changes over time, as when roaming users are involved.
%To date, we believe that the problem of controlling the computational complexity of NC has not yet received a satisfactory solution.

In this work, we achieve controlled-complexity NC by exploiting the concept of ``encoding window" to constrain the encoding and recombination process to a subset of the message symbols.
While a thorough description of the encoding window concept is provided in Section~\ref{sec:proposed_architecture}, here we review existing works based on this concept. 
In \cite{windowed2006} a window-based rateless coding mechanism is proposed to reduce the decoding complexity in a source-receiver scenario.
However, in \cite{windowed2006} there are no intermediate nodes that recombine packets and therefore the problem of preserving the degree distribution (one of the major contribution of our paper) is not addressed.
Therefore, as we detail in Section~\ref{sec:proposed_architecture}, the definition of the encoding window used in our paper is not the same as in \cite{windowed2006} due to the different goal of our work. 
Other related approaches based on the concept of encoding window can be found in \cite{luby2002patent, soro2009erasure}.
However, as in \cite{windowed2006}, both papers consider a simple source-receiver scenario and do not consider recombinations at the network nodes.
In detail, \cite{luby2002patent} proposes an LDPC scheme that aims at minimizing the memory accesses during encoding, but the properties of the code are shown for very large block sizes and windows, which would be an issue for multimedia applications.
In \cite{soro2009erasure}, a class of LDPC codes with a hybrid iterative/maximum likelihood decoding scheme is presented, where the generator matrix is designed to have a banded structure so to reduce the maximum likelihood decoding complexity.
Therefore, to the best of our knowledge, our work is the first to leverage the concept of encoding window to solve the problem of preserving the decoding complexity through the recombinations at the network nodes, which is the main novelty of our work.
 %With respect to our work, [soro2009] considers a source-receiver FEC scenario, which is rather different from the issue of controlling the decoding complexity through the recombinations at the network nodes, which is instead the main goal of our work.

%\section{Random Network Coding over GF(2)}
\section{Binary Network Coding}
\label{sec:rnc_over_gf2}

%In this section we describe the principles of random linear NC with rateless codes, i.e. NC over $GF(2)$.
In this section we describe the principles of binary NC, i.e. NC over $GF(2)$.
Then, we demonstrate how random packet recombinations at the nodes of a randomly connected graph, which is a case of particular interest for P2P communications, can drive the decoding complexity.

The source node holds a message $x$, i.e. a \emph{generation} of data, that has to be distributed to multiple network nodes.
The message is further subdivided in $N$ symbols $(x_0 , ... , x_{N-1})$, where $N$ is called \emph{generation size}.
Each time the opportunity to transmit a packet arises, the source produces a linear combination of the input symbols as $y = \sum_{i=0}^{N-1} g_i x_i$, where the sum operator represents the bit-wise XOR operator, indicated as $\oplus$ in the following.
%We assume that $g_i \in [0, 1]$ is randomly drawn from some distribution $\Omega^0$ that we call \emph{source distribution}.
The vector $g=(g_0, ..., g_{N-1})$, $g_i \in GF(2)$, is called \emph{encoding vector} and is created as follows.
Initially, the number $d$ of elements of $g$ that are equal to one, called the \emph{degree} of the encoded packet, is drawn according to a specific \emph{degree distribution} $\Omega$, i.e. such that $\Omega_d = \mathbb{P}\{\|g\|_0 = d\}$. 
Later, $d$ random elements of $g$ are set to one, while the remaining $N-d$ are set to zero. 
For Random Network Coding (RNC), $\Omega$ is the Binomial Distribution $\mathcal{B}(N, \frac{1}{2})$. 
If a rateless code is used at the source, the degree distribution of the code is used. 
%We assume that $g_i \in [0, 1]$ is randomly drawn from some distribution $\Omega^0$ that we call \emph{source distribution}.
%We consider symbols belonging to the Galois field $GF(2)$ and we denote as \emph{encoding vector} the binary array $g=(g_0, ..., g_{N-1})$, which has exactly $d$ non-zero elements.
Finally, the source transmits a packet $P (g, y)$ composed by the encoded payload $y$ plus the corresponding encoding vector $g$, so that the packet can be decoded at the network node~\cite{chou2003practical}.

The nodes of the network receive encoded packets and transmit random linear combinations thereof as follows.
Let us assume that a node has $k$ packets $(P^0, ..., P^{k-1})$ stored in its input buffer. 
In a RNC system, the recoder performs a linear combination of the payloads of the received packets as $y^r = \sum_{i=0}^{k-1} c_i y^i$ and their respective encoding vectors as $g^r = \sum_{i=0}^{k-1} c_i g^i$, where $c_i \in GF(2)$ and 
$\mathbb{P}\{c_i = 1\} = \frac{1}{2}$.  
%each element of $c=(c_0, ..., c_{k-1}) \in GF(2)$ is drawn at random.
The result of the recombination is packet $P^r (g^r, y^r)$ which is transmitted on the outgoing link of the node.
%The recombinations at the nodes increase the likelihood that the transmitted packet is linearly independent from all the packets previously collected by the receiver, thus increasing the useful network throughput.
%Whenever a node receives a packet that is linearly independent from all the previously received packets, such packet is said to be \emph{innovative}.
\\
Once a node has collected $N$ linearly independent packets, it recovers the original message by solving the system of equations corresponding to the received packets.
In the ideal case, all the packets received by the node are innovative and the generation is recovered after $N$ packets have been received.
In practice, not all the received packets are necessarily innovative due to the random encoding process at the source and the random recombination at the nodes.
Therefore, in practical cases the node needs to receive $N' \ge N$ packets to solve the system of equations.
%The encoding efficiency of a class of rateless codes can be measured in terms of the \emph{encoding overhead}, which is defined as $\epsilon=N'/N-1$.
%, where $N'\ge N$ is the average number of packets that the decoder must  receive to decode a generation of size $N$.
Once a node has received enough linearly independent packets, it solves the corresponding system of equations via some linear solving algorithm such as Gaussian Elimination (GE).
GE organizes the system of equations as a matrix of size $N' \times N$, where each row of the matrix is the encoding vector of one of the received packets.
The GE reduces the matrix to a triangular form via iterated XOR operations between the rows of the matrix.
The number of required XOR operations depends on the density of the matrix and hence on the average degree of the received packets as the rows of the matrix are the encoding vectors of the received packets.
However, the average degree of the received packets grows at each recombination and independently from the degree distribution exploited at the source as stated by the following proposition, whose proof is reported in appendix.

\begin{prop}
\label{prop:recombination_issue}
Random packet recombinations at the nodes of a random network alter the degree distribution selected at the source which converges to the Binomial Distribution $\mathcal{B}(N, \frac{1}{2})$ and the expected degree of the packets in the network tends to $\frac{N}{2}$.

\end{prop}
%It follows that random recombinations increase the density of the matrix to diagonalize and the number of XOR operations required to recover the message soars.
%Moreover, the packet recombination process is critical to control the computational complexity of the packet decoding process, as we show later on.

\begin{comment}
%The above proposition and its corollary yield two implications of paramount importance.
%First, lightweight decoding algorithms for rateless codes that rely on specific source distributions are made useless by the altered packet degree distribution.
%Second, the computational complexity of general purpose equation solving algorithms such as Gaussian Elimination grows with the degree of the packets.
%\textbf{XXX Attilio: Valerio suggerisce eventualmente di tagliare la dimostrazione e mettere un riferimento al paper Data obfuscation with network coding di A. Hessler, T. Kakumaru, H. Perrey, D. Westhoff, Computer Communications 35 (2012) 48–61}
\end{comment}

\begin{figure}
  \begin{center}
    \rotatebox{270}{\includegraphics[width=0.5\columnwidth]{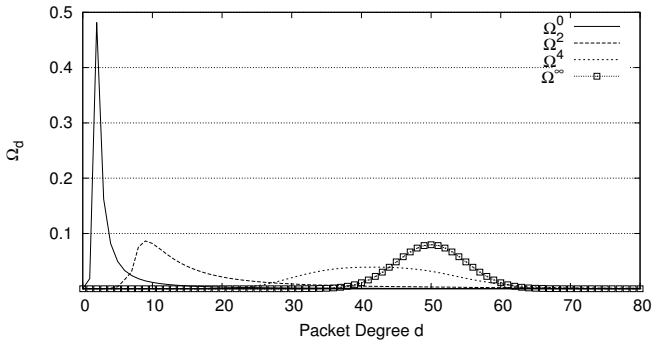}}
    \caption{Random recombinations alter the degree distributions of the packets. $\Omega^0$ is the RSD Distribution at the source, $\Omega^\infty$ is the asymptotic $\mathcal{B}(N, \frac{1}{2})$ distribution at the nodes.}
    \label{fig:rsd_ddp}
  \end{center}
\end{figure}

We exemplify how random recombinations at the nodes of a randomly connected network alter the packet degree distribution imposed at the source.
We assume that the source draws the degree of the encoded packets according to the Robust Soliton Distribution (RSD) used in LT codes~\cite{LT}.
%LT codes achieve a favorable tradeoff between encoding overhead and decoding complexity thanks to the joint design of the source distribution and the decoding algorithm.
%The RSD guarantees that some packets are encoded with degree $d=1$, i.e. some packets contain a non decoded symbol $x_i$, which enables the network nodes to decode the message with the lightweight Message Passing algorithm.
%The packets are decoded with a lightweight algorithm known as Message Passing (MP,) which is executed each time a packet $P$ with degree equal to 1 is received.
%MP reduces by one the degree of all the previously received packets that were encoded as a linear combination of $x_i$ performing an XOR with $x_i$.
%The MP algorithm iterates until it is not able to proceed further with the decoding process or until it solves the system of linear equations.
%The RSD guarantees that some packets are encoded with degree equal to one and thus the generation is likely to be decoded by MP.
Figure~\ref{fig:rsd_ddp} illustrates how the packet degree distribution in the network evolves as a function of the number of recombinations for a generation of $N$=100 symbols.
The curve $\Omega^0$ show the RSD distribution imposed by the source node, curves $\Omega^2$ and $\Omega^4$ show the degree distribution in the network after 2 and 4 recombinations respectively and curve $\Omega^\infty$ is the degree distribution after a number of recombinations that tends to infinite.
After just two recombinations, the packet degree distribution has considerably changed with respect to the original source distribution.
Finally, as the number of recombinations further increases, the packet degree distribution converges to the Binomial Distribution $\mathcal{B}(100, \frac{1}{2})$ and the average degree of the packets in the network tends to $\frac{N}{2}$=50 as stated in Proposition~\ref{prop:recombination_issue}.
%The MP algorithm is unsuited to decode the generation due to the lack of packets with degree equal to one and the more complex GE has to be used in its place.
%Moreover, as the computational complexity of GE depends on the degree of the received packets, that yields to a rather complex decoding process as explained before.

\section{Band Codes}
\label{sec:proposed_architecture}

\begin{figure}[h]
  \begin{center}
	\includegraphics[width=0.7\columnwidth]{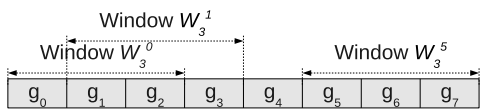}
	\caption{Encoding vector for a generation of size $N=8$ and three encoding windows of size $W=3$.}
	\label{fig:encoding_window}
  \end{center}
%    \vspace{-10mm}
\end{figure}

In this section we describe the algorithms and the related properties that enable BC to preserve the packet degree distribution through the recombinations at the network nodes.
We prove that BC packet degree distribution is preserved by design under an arbitrary number of 
recombinations in the network.  
This property holds for an arbitrary network, although the end-to-end coding efficiency depends 
on the overlay topology as well.  
The definition of BC covers not only the packet encoding process at the source node, but also the packet recombination and the decoding processes at the network nodes, and is based on the concept of \emph{encoding window}.
%, a class of codes that enable to recombine packets at the network nodes preserving the packet degree distribution.
%With respect to other classes of rateless codes, the definition of BC covers not only the packet encoding process at the source node, but also the packet recombination process at the network nodes.
%BC achieve controlled decoding complexity thanks to the joint design of the packet encoding process at the source node and the packet recombination process at the network nodes.

Let us consider a generation composed of $N$ symbols $(x_0, ..., x_{N-1})$ and a generic encoding vector $g = (g_0, ..., g_{N-1})$.
We define encoding window a subset of the encoding vector $(g_f, ..., g_l)$ where $f$ and $l$ are, respectively, the \emph{leading edge} and the \emph{trailing edge} of the window.
The elements of the encoding vector that do not belong to the encoding window are equal to zero.%\textbf{XXX Attilio: questo è vero, pero mi sembra che anticipi un po' cio che poi diciamo sotto nell ncoding process.}
We define the size of the encoding window as $W = l-f+1$; for a generation of size $N$, there are $N-W+1$ possible encoding windows of size $W$.
We indicate as $\mathcal{W}^{f}_{W}$ the encoding window that has size $W$ and leading edge $f$.
Figure~\ref{fig:encoding_window} shows the encoding vector for a generation of size $N=8$ and three of the $N-W+1=6$ possible encoding windows $(\mathcal{W}_3^0, \dots, \mathcal{W}_3^5)$ of size $W=3$.
Let us now consider two encoding windows $\mathcal{W}_{W^1}^{f^1}$ and $\mathcal{W}_{W^2}^{f^2}$ such that $f^2 \geq f^1$:  we say that $\mathcal{W}_{W^1}^{f^1}$ and $\mathcal{W}_{W^2}^{f^2}$ \emph{overlap} if $f^2 \leq l^1$, i.e. if $f^1 \leq f^2 \leq l^1$.
For example, in Figure~\ref{fig:encoding_window}, window $\mathcal{W}_3^0$ overlaps with $\mathcal{W}_3^1$ and $\mathcal{W}_3^2$.
%In windowed codes the encoding window can wrap around the boundaries of the generation, i.e. it is possible that $l < f$.
%The reason why in BC the encoding window is not allowed to wrap is related to the decoding and recombination process of BC and will be clarified later on.
A key difference from~\cite{windowed2006} is that in our work the encoding window is not allowed to ``wrap around" the encoding vector boundaries, as the wrap would be an issue with the recombinations.
So, in the design of both the encoding and recombination algorithms of BC, we do not allow the window to wrap around and hence we impose that $0 \leq f < l \leq N-1$.
The reasons behind this and the other differences between the design of BC and other families of codes will be detailed in the rest of this Section.
Equipped with the concept of encoding window, we proceed to describe the process to encode packets at the source.

\subsection{Packet Encoding at the Source}
\label{sec:proposed_architecture_encoding}

In the following, we describe the encoding process at the source node for a generation of size $N$ and an encoding window of size $W$.
Every time a new packet $P(g,y)$ has to be encoded, the source draws the position of the leading edge $f$ of the encoding window according to the distribution:

\begin{equation*}
\begin{array}{c}
HD_{N,W}(f) = \left\{
\begin{array}{cl}
\frac{W+1}{2N} & \textrm{if~~~~$f=0$ or $f=N - W$} \\[+2pt]
\frac{1}{N} & \textrm{if~~~~$0 < f < N - W $} \\
\end{array}
\right.
\end{array}
\end{equation*}

\noindent
As in BC the encoding window is not allowed to ``wrap around" as in~\cite{windowed2006}, not all symbols $x_i$ are contained by the same number of encoding windows.
If $f$ was drawn with uniform probability, some symbols would appear in the coded packets with very low probability, impairing the encoding efficiency.
The $HD_{N,W}(f)$ guarantees that each symbol $x_i$ is selected for encoding with similar probability, which in turn improves the decoding efficiency.\\
%Then, the elements $g_i$ of the encoding vector are set to one with probability $p = \frac{1}{2}$ for $i \in [f,l]$ and $p=0$ otherwise.
Then, the elements of the encoding vector $g$ are set to zero outside of the encoding window and to one with probability $\frac{1}{2}$ inside the encoding window, i.e.
\begin {equation}
\label{eqn:encoding_scheme}
\ \mathbb{P}\{g_i = 1\} = \left\{
\begin{array}{cl}
\frac{1}{2} & \textrm{if~~~~$f \le i \le l$,} \\
0 & \textrm{otherwise}. \\
\end{array}
\right.
\end {equation}
\noindent
Let $s$ and $t$ be the positions of the first and the last non-zero elements of $g$ , i.e. $g_s = g_t = 1$ and $g_i = 0$ for any $i < s$ and for any $i > t$: we call $g_s$ and $g_t$, respectively, the \emph{leading one} and the \emph{trailing one} of the encoding vector.
Finally, the source combines the original symbols as $y = \sum_{i=0}^{N-1} g_i x_i$ and transmits the encoded packet $P(g,y)$ to the network.

In the following, we say that a packet $P(g,y)$ is a \emph{Band Packet} $BP(N, W)$ with leading edge $f$ if $g_i = 0$ for $i \notin [f, l]$ and $\mathbb{P}\{g_i = 1\} = \frac{1}{2}$ for $i \in [f, l]$.
The degree distribution of a $BP$ follows the Binomial Distribution $\mathcal{B}(W, \frac{1}{2})$ and the expected packet degree is equal to $\frac{W}{2}$.
%In fact, $\Omega^0_i = \mathbb{P}\{\|g\|_0 = i\}$ corresponds to the random variable that counts the number of times that $g_i = 1$ for $i \in [f,l]$, thus $\Omega^0 \sim \mathcal{B}(W, \frac{1}{2})$.
%The packet encoding procedure described above is such that any packet encoded by the source node is a
It follows that packets encoded according to (\ref{eqn:encoding_scheme}) are $BP(N,W)$ and the following property holds.

\begin{prop}
\label{prop:xor}
Let $P^1(g^1, y^1)$ and $P^2 (g^2, y^2)$ be two packets that are, respectively, $BP(N, W^1)$ and $BP(N, W^2)$. If the respective encoding windows $\mathcal{W}_{W^1}^{f^1}$ and $\mathcal{W}_{W^2}^{f^2}$ overlap, then packet $P^r (g^r, y^r) = P^1 \oplus P^2$ is a $BP(N, W^r)$ whose encoding window size is $W^r \le W^1 + W^2$.
\end{prop}
\begin{corol}
\label{cor:xor}
Under the hypothesis of Prop.~\ref{prop:xor}, if $W^1 = W^2 = W$ and $f^1= f^2$ then $W^r = W$.
\end{corol}
\noindent
It follows from the definition of BP that the degree distribution of the recombined packet $P^r$ follows the Binomial Distribution $\mathcal{B}(W^r, \frac{1}{2})$. 
%Moreover, under the condition $W^1 = W^2$ and $f^1= f^2$, such property enables to preserve the packet degree distribution, i.e. $W^r = W^1$ as shown in Section~\ref{sec:proposed_architecture_recombination}.
%\end{comment}

\subsection{Packet Decoding at the Network Nodes}
\label{sec:proposed_architecture_decoding}

For the decoding process, we assume that the nodes of the network receive packets $P(g,y)$ that are $BP(N,W)$. 
While this assumption is obviously true for packets received from the source, in the next section we will prove that the assumption holds also for packets received from the network nodes. 
The received packets are decoded using an ad-hoc modified version of early-decoding Gaussian Elimination, that we called ``Swap Gaussian Elimination" (SGE).
Similarly to the early-decoding GE algorithm, SGE organizes the system of received equations into a triangular matrix that avoids the triangularization process typical of standard GE.
%Moreover, SGE performs a preliminary recombination of the received packets which, increases the probability to produce innovative packets when combining the rows of $G$ as detailed in Section~\ref{sec:proposed_architecture_recombination}.
SGE solves a system of $N$ linear equations $G X = Y$, where $G$ is the $N \times N$ matrix that stores linear combinations of the encoding vectors of the received packets. % as illustrated in Figure~\ref{fig:band_G} for the toy case $N=8$.
In the following, we use the notation $G_i$ to indicate the $i$-th row of $G$ and we use the notation $G_{i,j}$ to indicate the element of $G$ at row $i$, column $j$.
When all the elements of $G_i$ are equal to zero, we say that the row $i$-th is empty and we write $G_i = \emptyset$.
%Similarly, if all the rows of $G$ are empty, we define $G$ as empty and we write $G=\emptyset$.
$Y$ is the $N \times 1$ vector that stores the combinations of the payloads $y$ of the encoded packets received by the node and $X$ is the $N \times 1$ vector that contains the symbols $x_i$ to recover.
For the sake of simplicity, we describe the operations on the $G$ matrix and the received encoding vectors $g$ and we omit the equivalent operations on $y$ and $Y$.

\begin{figure}[h]
  \begin{center}
    \includegraphics[width=0.65\columnwidth]{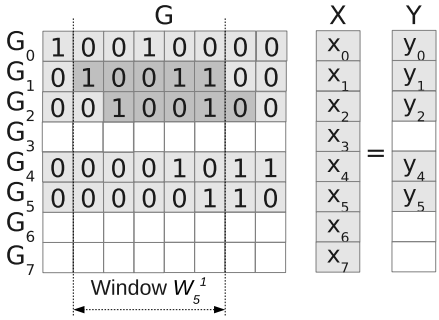}
	\caption{Example of matrix G for a generation of N=8 symbols and an encoding window of size W=5. G The matrix is upper-triangular and its rank is equal to the number of linearly independent packets received by the node (5 in the example). Rows 1 and 2 are suitable for recombination given an encoding window of leading edge $f^r=1$}
	\label{fig:band_G}
  \end{center}
%  \vspace{-5mm}
\end{figure}

The SGE algorithm operates in two stages, triangularization and diagonalization, as follows.
\\
%The triangularization stage arranges the equations corresponding to the received packets into $G$, which is initially empty.
%The received equations are processed so that $G$ is built upper-triangular, which avoids the final triangularization typical of standard GE and distributes the computational load of the decoding process over time.
The triangularization stage is formalized as Algorithm~\ref{alg:ofg_stage_one} and it is executed every time a new packet $P (g,y)$ is received.
Let $g_s$ be the leading one of g, i.e. the first non-zero element of $g$.
%Depending on whether $G_s$ is empty or not, the algorithm operates as follows.
If $G_s$ is empty, $g$ is inserted in the $s$-th row of $G$ (line 5) and the algorithm ends.
Otherwise $g$ and $G_s$ are swapped (line 8) to refresh the rows of $G$ between two consecutive transmission opportunities of the node.
This swap increases the probability to recombine an innovative packet if few rows of $G$ are suitable to be recombined as explained in Section~\ref{sec:proposed_architecture_recombination}.
The swap does not further increase the computational complexity as it is performed by swapping two pointers in memory.
At this point, if $g$ is identical to $G_s$, $P$ is not innovative and the algorithm ends.
Otherwise a XOR between $g$ and $G_s$ is executed and the algorithm iterates.
Figure~\ref{fig:band_G} shows matrix $G$ after 5 independent packets have been received by the node for a toy case where $N$=8 and $W$=5.

\begin{algorithm}[h!]
  \caption{SGE, Triangularization}
  \label{alg:ofg_stage_one}
  \begin{algorithmic}[1]
  \STATE \textbf{receive}~$P(g)$.
  \WHILE{true}
  \STATE $s \gets$ position of leading one of $g$ %equal to 1.
    \IF {$G_s = \emptyset$}
      \STATE $G_s \gets g$
      \STATE \textbf{end}
    \ELSE
%      \STATE \textbf{if} $(deg(g) < deg (G[f]))$
%      \STATE \textbf{if} $\sum_{i=0}^{N-1} g_i < \sum_{i=0}^{N-1} G_{s,i}$ \textbf{XXX qui usare $\|g\|$ ?}
%      \IF{$(deg(g) < deg (G[f]))$}
      \STATE swap $G_s$ and $g$
%      \ENDIF
%      \IF{$g = G[f]$}
      \STATE \textbf{if} $g = G_s$
        \STATE ~~~~\textbf{end}
%      \ENDIF      
      \STATE $g \gets g \oplus G_s$
    \ENDIF
  \ENDWHILE
  \end{algorithmic}
\end{algorithm}
\begin{comment}
and is formalized as Algorithm~\ref{alg:ofg_stage_two}.

\begin{algorithm}[h!]
  \caption{On-the-Fly GE (OFG), Second Stage}
  \label{alg:ofg_stage_two}
  \begin{algorithmic}[1]
    \FOR {$i = N -1 \to 0$}
      \WHILE {degree of $G_i > 1$}
        \STATE $t \gets$ position of the $2$-nd element of $G_i$ equal to 1.
        \STATE $G_i \gets G_i \oplus G_t$
      \ENDWHILE
    \ENDFOR
  \end{algorithmic}
\end{algorithm}
\end{comment}

%Figure~\ref{fig:band_G} shows an example of matrix $G$ for the toy case $N=8$ and $W=5$ after 5 independent packets have been received by the node.
Under the assumption that node receives packets $P(g,y)$ that are $BP(N,W)$, the following proposition about the structure of matrix $G$ holds:
\begin{prop}
\label{prop:G_rows_are_BC}
%If the SGE processes packets $P(g,y)$ that are $BP(N,W)$, then $G$ is an upper triangular band matrix (i.e., $G_{i,j} = 0$ if $j<i$ or $j>i+W-1$) and each pair $(G_i,Y_i)$ is a $BP(N, W)$.
If the SGE processes packets $P(g,y)$ that are $BP(N,W)$, then $G$ is an upper triangular band matrix (i.e., $G_{i,j} = 0$ if $j<i$ or $j>i+W-1$) and $G_i$ is the encoding vector of packet $P(G_i, Y_i)$ that is a $BP(N, W)$.
\begin{IEEEproof}
The proof is built by mathematical induction on the number of received packets $k$.\\
\textit{Basis:} $G$ is empty when the first packet $P(g,y)$ is received and $g$ is inserted in $G_s$, where $s$ is the position of the leading one of $g$.
$G_s$ is equal to $g$ and since $P$ is a $BP(N, W)$, the proposition is proved.\\
\textit{Inductive step:} Let us assume that the proposition holds for the first $k$ received packets.
The $(k+1)$-th Packet $P(g,y)$ is then received, and we have two cases to consider.
if $G_s$ is empty, then $g$ is inserted in $G_s$ and the proposition holds for the same reason illustrated in the Basis step.
Otherwise, a swap between $G_s$ and $g$ happens and the proposition still holds for the same reason.
Then, the algorithm computes $g = g \oplus G_s$: $g$ and $G_s$ are the encoding vectors of two $BP(N, W)$ with the same leading edge, %and thus the proposition still holds due to Proposition~\ref{prop:xor}.
hence the updated $g$ is a $BP(N, W)$ due to Cor.~\ref{cor:xor}, and the proposition still holds. 
\end{IEEEproof}
\end{prop}
\noindent
%Such proposition enables to calculate the expected number of elements of $G$ equal to 1 as a function of $W$ and is exploited in Section~\ref{sec:complexity_model} to model the computational complexity of OFG.
%Furthermore, we exploit such proposition in the following section where we describe an algorithm to recombine the received equations at the nodes.
Once $N$ linearly independent packets have been received, the rank of $G$ is equal to $N$.
At this point, the matrix is diagonalized through iterated XOR operations among its rows and at the end the vector $X$ contains the recovered symbols $x_i$.

\subsection{Packet Recombination at the Network Nodes}
\label{sec:proposed_architecture_recombination}

Each time a transmission opportunity arises, a network node transmits a linear combination of a subset of rows of the $G$ matrix.
The rows of $G$ form in fact an (incomplete) basis for the system of linear equations that the nodes must solve to recover the original symbols, so the linear combination thereof is a linear combination of the original input symbols.
Moreover, proposition~\ref{prop:G_rows_are_BC} states that rows of $G$ are $BP(N, W)$, so the linear combinations thereof is a $BP(N,W)$ given the recombination process detailed below formalized as Algorithm~\ref{alg:recombination}. %and takes as input the size $W$ of the encoding window of the packet $P^r(g^r, y^r)$ to transmit.
First, the algorithm draws the encoding window leading edge $f^r$ from the distribution $HD_{N,W}$ and computes the relative trailing edge $l^r = f^r+W-1$.
Let $R = \{G_{q_{0}}, \dots, G_{q_{|R|-1}}\}$ be the set of rows of $G$ for which $s_{q_i} \geq f^r$ and $t_{q_i} \leq l^r$, where $s_{q_i}$ and $t_{q_i}$ are the position of the leading and trailing one of $G_{q_i}$.
If $R = \emptyset$, i.e. there are no suitable rows for recombination as it may happen when just a few packets have been received, $f^r$ is drawn again.
The encoding vector $g^r$ is now calculated as a linear combination of the elements of $R$, i.e. $g^r = \sum_{i=0}^{|R|-1} c_i G_{q_i}$ where $c_i$ is drawn at random so that $\mathbb{P}\{c_i = 1\} = \frac{1}{2}$.
Similarly, the payload $y^r$ is calculated as $y^r = \sum_{i=0}^{|R|-1} c_i Y_{q_i}$.
Due to Prop.~\ref{prop:xor}, the recombined packet $P^r$ calculated by Algorithm~\ref{alg:recombination} is a $BP(N, W)$, hence the recombination process preserves the packet degree distribution selected at the source. \\
Figure~\ref{fig:band_G} shows an example of the recombination process where $f^r=1$ and $l^r=5$.
In this example, rows $G_1$ and $G_2$ are suitable for recombination.
$G_1$ is suitable for recombination because its leading and trailing edges are equal to $f^r$ and $l^r$, respectively.
While the trailing edge of $G_2$ is greater than $l^r$, its trailing one is equal to $l^r$, so $G_2$ is still suitable for recombination.
As this example illustrates, a few rows of $G$ only may be suitable for recombination due to the constraint on the $G$ rows, especially during the initial stages of the transmission.
However, the probability that a packet is is innovative depends, among others, on the number of independent packets already collected by the receiver.
The decoding process proceeds almost in parallel at all the network nodes, i.e. all the nodes have received a similar number of packets at a given moment.
Therefore, even in the case where the transmitter has few rows to recombine, also the receiver has received few rows yet, which increases the probability to encode an innovative packet.

\begin{algorithm}
  \caption{Packet Recombination} %(Given $W$)}
  \label{alg:recombination}
  \begin{algorithmic}[1]
  \STATE $R = \emptyset$
  \WHILE{$R = \emptyset$}
     \STATE Draw $f^r$ as $HD_{N,W}(f^r)$
     \STATE $R \gets $ set of rows of $G$ s.t. $s_{q_i} \geq f^r$ and $t_{q_i} \leq l^r$
  \ENDWHILE 
  \STATE $g^r = \sum_{i=0}^{|R|-1} c_i G_{q_i}$;~~~$y^r = \sum_{i=0}^{|R|-1} c_i Y_{q_i}$
  \STATE \textbf{transmit} $P^r (g^r, y^r)$
  \end{algorithmic}
\end{algorithm}

\subsection{Modeling the Decoding Complexity of BC}
\label{sec:complexity_model}

%In this section we model the decoding complexity of the OFG algorithm for decoding a generation, i.e. the average number of required operations to recover the original message, under the assumption that the algorithm processes packets that are $BP(N,W)$.
%We define the \emph{decoding complexity} ($C_D$) of BC as the average number of XOR operations required to recover a generation with the SGE algorithm under the assumption that a node receives packets that are $BP(N,W)$.
We define the \emph{decoding complexity} ($C_D$) of BC as the computational complexity of the SGE algorithm under the assumption that a network node receives packets that are $BP(N,W)$.
In detail, we are interested in calculating the average number of XOR operations between rows required to recover a generation as a function of the generation size $N$ and the encoding window size $W$. 
The decoding complexity can be calculated as the sum of the decoding complexities of the triangularization and the diagonalization stages. 
%The analysis of the OFG shows that the XOR operator is the major consumers of processor cycles, as each XOR requires access large amounts of memory.
%On the other hand, operations such as a swap between rows can be implemented exchanging two pointers in memory, thus their contribution to the overall processor load is negligible.
%For this reason, in this section we focus on modeling the average number of XOR operations required by OFG to decode a generation.
\\
The decoding complexity of the triangularization stage is defined as $C_D^{(t)}$ and increments by one unit every time the XOR at line 11 of Algorithm~\ref{alg:ofg_stage_one} is executed.
%We assume that all the packets received by the node are innovative, i.e. the rank of $G$ is equal to $k$ after the $k$-th packet is received and each collision does result in a XOR.
%We first estimate the maximum number of collisions that may happen while decoding the $(k+1)$-th received packet $P (g, y)$.
%When $P$ is received, the maximum number of rows of $G$ with whom $P$ can collide is $k$.
We conduct a worst case analysis for the case where the rank of $G$ is equal to $k$ after the $k$-th packet is received, i.e. assuming that all the packets received by the node are innovative. 
%This model turns out to be accurate average case for $\epsilon > 0$, as shown in Sect.~\ref{sec:rnc_experiments}. 
However, this model turns out to be accurate also for the case where $rank(G) < k$, as shown in Section~\ref{sec:rnc_experiments}. 
When the $(k+1)$-th packet $P (g, y)$ is received, the maximum number of rows of $G$ with whom $P$ can collide is $k$. 
On the other hand, since $g$ has average degree $\frac{W}{2}$, each row will collide with a probability of $\frac{W}{2N}$.
Therefore, the average number of collisions that happen when the $(k+1)$-th packet is received is equal to $\frac{k}{N} \frac{W}{2}$.
So, the decoding complexity of the triangularization stage is

\begin{equation*}
C_{D}^{(t)} = \sum_{k=1}^{N-1} \frac{kW}{2N} = \frac{W(N-2)}{4}.
% \label{eqn:cost_decoding_one}
\end{equation*}
\noindent
The decoding complexity of the  diagonalization stage is defined as $C_D^{(d)}$ and increments by one unit every time an XOR is executed.
The number of required XORs depends on the density of the $G$ matrix, that is on the number of elements of $G$ that are equal to 1 except for those on the diagonal.
$G$ is an upper triangular band matrix with band width $W$, hence it has at most $N^2 - \frac{(N-1)^2}{2} - \frac{(N-W)^2}{2}$ non-zero elements. 
The $N$ elements on the diagonal are not involved in the backward substitution so they shall not be accounted, while the remaining elements of $G$ are non-zero with probability $\frac{1}{2}$.
%Due to proposition~\ref{prop:G_rows_are_BC}, the expected degree of the first $N-W$ rows of $G$ is equal to $\frac{W}{2}$, thus $\frac{W}{2} - 1$ XOR operations are required to diagonalize each of the first $N-W$ rows of $G$.
%The expected degree of the remaining $W$ rows of $G$ is equal to $\frac{W}{4}$, thus $\frac{W}{4} - 1$ XOR operations are required to diagonalize each of the last $W$ rows of $G$.
%Hence, as each row of $G$ has $d_G - 1$ non null coefficients excluding  that on the diagonal, the average number of XORs executed to diagonalize $G$ is equal to $(d_G - 1) N \simeq \frac{W}{4} N$.
So the decoding complexity of the  diagonalization stage is

\ifDRAFTMODE
\begin{equation*}
	%C_{D}^{(2)} = \frac{(W-2)N}{4}.
	%C_{D}^{(d)} = (N-W)(\frac{W}{2}-1) + W(\frac{W}{4}-1) = \frac{2NW - W^2 -4N}{4}.
	C_{D}^{(d)} = \frac{1}{2} \left( N^2 - \frac{(N-1)^2}{2} - \frac{(N-W)^2}{2} - N \right) = \frac{2NW - W^2 -1}{4}.
	% \label{eqn:cost_decoding_two}
\end{equation*}
\else
	\[\begin{array}{r c l}
	C_{D}^{(d)} & = & \frac{1}{2} \left( N^2 - \frac{(N-1)^2}{2} - \frac{(N-W)^2}{2} - N \right)\\
	& = & \frac{2NW - W^2 -1}{4}.
	\end{array}
	\]
\fi
\noindent
Finally, the decoding complexity of BC is equal to

\begin{equation}
%C_D = C_{D}^{(1)} + C_{D}^{(2)} = \frac{WN -W -N}{2} \simeq \frac{WN}{2}.
C_D = C_{D}^{(t)} + C_{D}^{(d)} = \frac{3NW - W^2 -2W -1}{4}.
\label{eqn:cost_decoding}
\end{equation}
\noindent
The model shows that the decoding complexity of BC is $O(NW)$, i.e. it is in the order of $O(N^2)$ as for traditional RNC with GE.
However, the model shows that the actual decoding complexity of BC is a function of $W$, which is what enables to control the number of processor cycles required to recover the generation and makes NC practically feasible on mobile devices.
\section{P2P Video Streaming with Band Codes}
\label{sec:protocol}

In this section we describe the protocol for P2P video streaming that we use to analyze the performance of Band Codes.
The protocol we present is an extension of the protocol previously described in~\cite{fiandrotti2011complexity} and the source code is made freely available under GPL license.\footnote{http://www1.tlc.polito.it/oldsite/sas-ipl/torostream/}
% enhanced the protocol under many aspects such as the overlay management scheme and the feedback mechanism strategy, making it competitive with state-of-the-art such as $R^2$~\cite{wang2007r2}.
While we focus on video streaming, the mechanisms we describe are also suitable for other forms of cooperative data dissemination such as file sharing.

\subsection{Overlay Setup and Management}

We model the network as a graph $G(V,E)$ where each vertex $V = \{ \mathcal{N}_0, ..., \mathcal{N}_{|V-1|} \}$ of the graph is a node of the network.
A central tracker organizes the network nodes into a randomly connected graph where two vertices connected by an edge are known as \emph{peers}.
The unstructured mesh topology requires simple overlay management policies and offers increased resilience to peer churning.
%The recombinations at the nodes make in fact not strictly necessary to organizing the nodes into an acyclic graph, which results in a simpler overlay management policy.
The procedure for a node $\mathcal{N}^{i}$ to join the overlay is as follows.
The node first contacts the tracker and the tracker adds the node to a master list.
The tracker replies with a list of addresses of the nodes already in the overlay and the node starts a separate handshake with each address in the list.
\begin{comment}
For example, let us assume that node $\mathcal{N}_{i}$ starts the handshake with peer $\mathcal{N}_{j}$.
Node $\mathcal{N}_{i}$ transmits an integer value $W^i \in [1, N]$ to $\mathcal{N}_{j}$, where $W^i$ is the size of the encoding window of the packets $BP(N,W)$ that node $\mathcal{N}_{i}$ wishes to receive.
Similarly, node $\mathcal{N}_{j}$ transmits the integer value $W^j$ to node $\mathcal{N}_{i}$.
Each node stores the received $W$, which is exploited by the packet scheduling scheme described in the following to control the decoding complexity of the network nodes.
\end{comment}
Upon handshake completion, the two nodes become peers and start to exchange video packets without any exchange of buffermaps thanks to the embedded feedback mechanism described in the following.
Peers periodically exchange keep-alive messages to detect failures of the network.
If a node does not receive any message from a peer for too long, the peer is assumed to be unreachable and the relationship between peers is terminated with an appropriate message.

\subsection{Decoding Maps for Embedded Feedback}

\begin{figure}[h]
  \begin{center}
	\rotatebox{0}{\includegraphics[width=0.7\columnwidth]{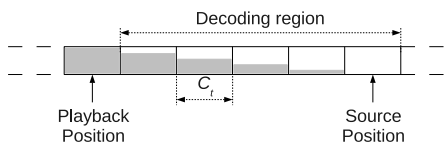}}
	\caption{The video stream as a sequence of generations as seen by a peer node. At any moment, a peer is interested in decoding the generations within its decoding region.}
	\label{fig:buffering_scheme}
  \end{center}
  \vspace{-5mm}
\end{figure}

Figure~\ref{fig:buffering_scheme} depicts the video stream organized as a sequence of independent generations.
A generation corresponds to one or multiple self-decodable units of video, e.g. Groups of Pictures (GOPs), and the generations have identical playback duration $C_t$.
While generations of different size are supported by the protocol, in the following we assume that the video is encoded at constant bitrate and thus all generations have also approximately equal size.
Every $C_t$ seconds, the source node fetches one generation of video from the source, e.g. a pre-encoded sequence or a live camera.
For the following $C_t$ seconds, the source encodes and distributes packets only for that generation, which we define as the \emph{source position} in the video stream.
Before starting the playback, a peer node buffers $t_b$ seconds of video, which correspond to $t_b / C_t$ generations.
After $t_b$ seconds, the node sends the first decoded generation to the video player and updates its playback position, and so on.
The generation of video currently reproduced by a node is called the \emph{playback position} of the node.
%Every $C_t$ seconds, the node updates its playback position and the playback continues uninterrupted for the rest of the duration of the video.
If a generation of video is not recovered before its playback deadline, the playback freezes and the quality of the video experience degrades.
%The playback position of a peer lags $t_b / C_t$ generations behind the seeding position and $t_b$ is the same for all the peers, which are all synchronized on the same playback position.

The set of generations encompassed between the playback position of a node and the source position in the video stream is referred to as \emph{decoding region} of the node.
We define as \emph{decoding map} the array of binary variables that describes the status (decoded or not) of the generations within the decoding region of a node.
The decoding region of a node corresponds to few generations of video, so a decoding map can be represented with few bits.
For example, if $C_t = 1s$ and $t_b = 5 s$, we have that the decoding map required $\frac{t_b}{C_t}$ = 5 bits.
%During the initial handshake, each pair of nodes exchange decoding maps for the respective decoding regions.
A decoding map is embedded in each handshake message, enabling two nodes to start transmitting immediately after the handshake.
Every time a node decodes a generation, it signals the event to all its peers with a specific \textit{stop} message.
When a node receives a stop message from one of its peers, the node updates the decoding map relative to the peer.
Stop messages may however be lost, thus each node also embeds its decoding map into each packet sent to its peers.
Embedding a decoding map in every transmitted packet increases the likelihood of timely feedback at the cost of a negligible increase in signaling bandwidth (e.g., given a decoding map of 5 bits and encoded video packets of 1250 bytes, the signaling overhead increase is around 0.05\%.)
%This mechanism enables the nodes to be timely updated about the decoding status of their peers and is exploited in the following to design a push-based packet scheduler.

\subsection{Packet Scheduling}

Each node of the network is periodically given an opportunity to transmit a packet to any of its peers, and at each transmission opportunity the packet scheduler is invoked.
%In our design, the packet scheduler does not just decide how to allocate the node output bandwidth, but also controls the computational complexity of its peer network nodes.
The scheduler selects which peer to address with a round-robin policy, so that the output bandwidth of the node is allocated in a fair way.
%In our exemplified scenario with just two nodes in the overlay, the scheduler always selects node $\mathcal{N}^i$.
Let us assume that node $\mathcal{N}^j$ has the opportunity to transmit a packet and the scheduler selects peer $\mathcal{N}^i$ as recipient for the transmission.
The scheduler checks the stored decoding map relative to $\mathcal{N}^i$ for those generations that have not been decoded yet and one of them is drawn at random with geometric probability.
%If the peer has already decoded all the generations in its decoding region, the scheduler selects another peer.
%The generation that is closer to the playout deadline is selected for transmission, with an earliest-decoding-deadline policy.
At this point, the scheduler executes Algorithm~\ref{alg:recombination}, which recombines some of the rows of the $G$ matrix for the selected generation and produces a packet $P^r$ that is a $BP(N,W)$. 
%The encoding window size $W^i$ that was received from node $\mathcal{N}^i$ during the handshake is passed to Algorithm~\ref{alg:recombination} as first parameter, while the number of packets to recombine $N_r$ is passed as second parameter.
%Algorithm~\ref{alg:recombination} recombines some of the received equations and produces a packet $P^r$ that is a $BP(N,W^i)$.
The scheduler finally transmits the recombined packet to $\mathcal{N}^i$ and waits for the next transmission opportunity.

\section{Experimental Results}
\label{sec:rnc_experiments}

In this section, we first experiment in a controlled conditions testbed to evaluate BC in terms of encoding efficiency, decoding complexity and energy consumption both in a simple end-to-end scenario and in a more complex network scenario with recombinations at the nodes.
Then, we experiment on the Internet to verify which video quality BC enable in real networks.

\subsection {End-to-End Performance of BC}

\begin{figure}[h]
  \begin{center}
  	\subfigure{
	\rotatebox{270} {
	  \includegraphics[width=0.315\columnwidth]{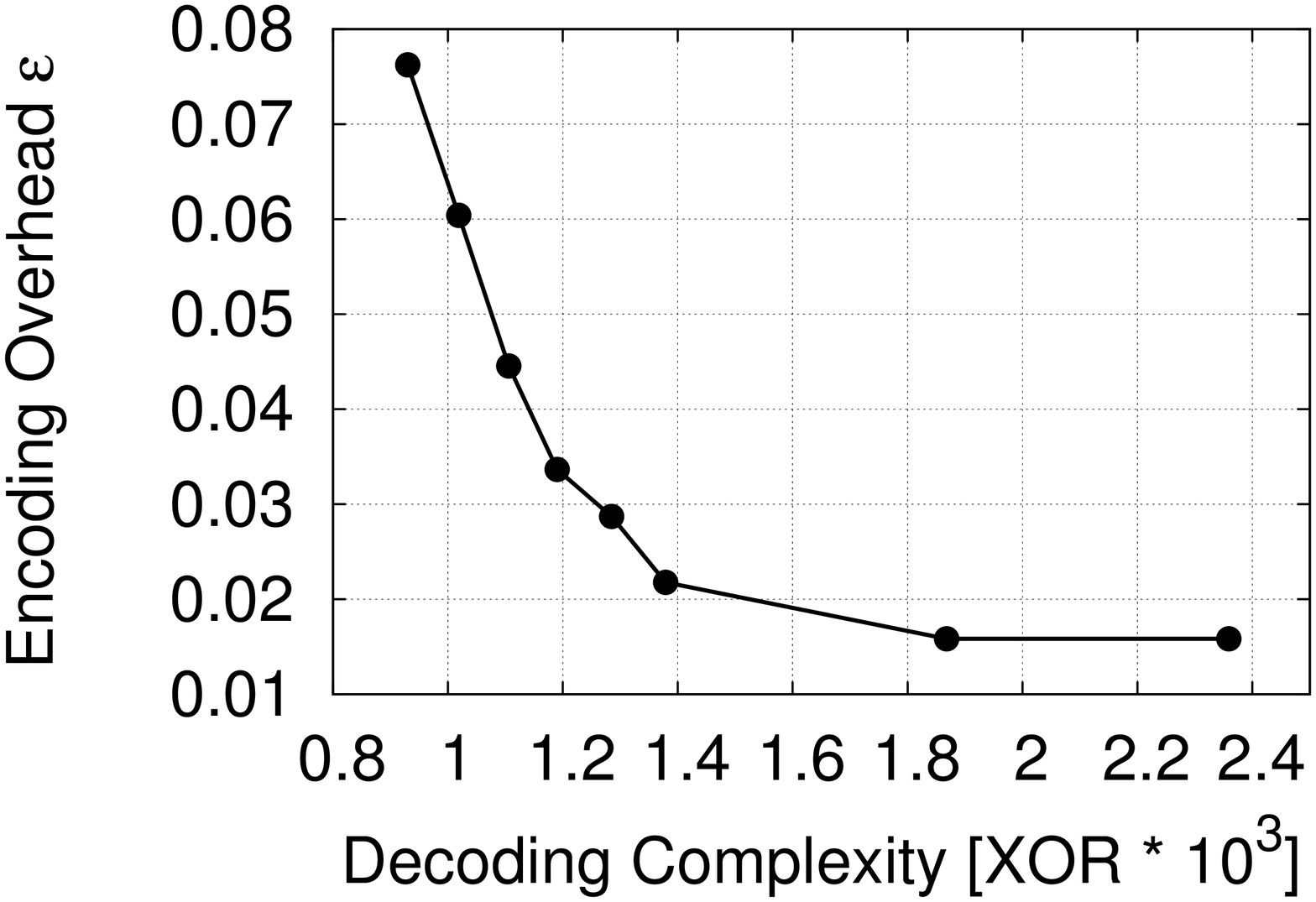}
	}
	}
  	\subfigure{
	\rotatebox{270} {
	  \includegraphics[width=0.315\columnwidth]{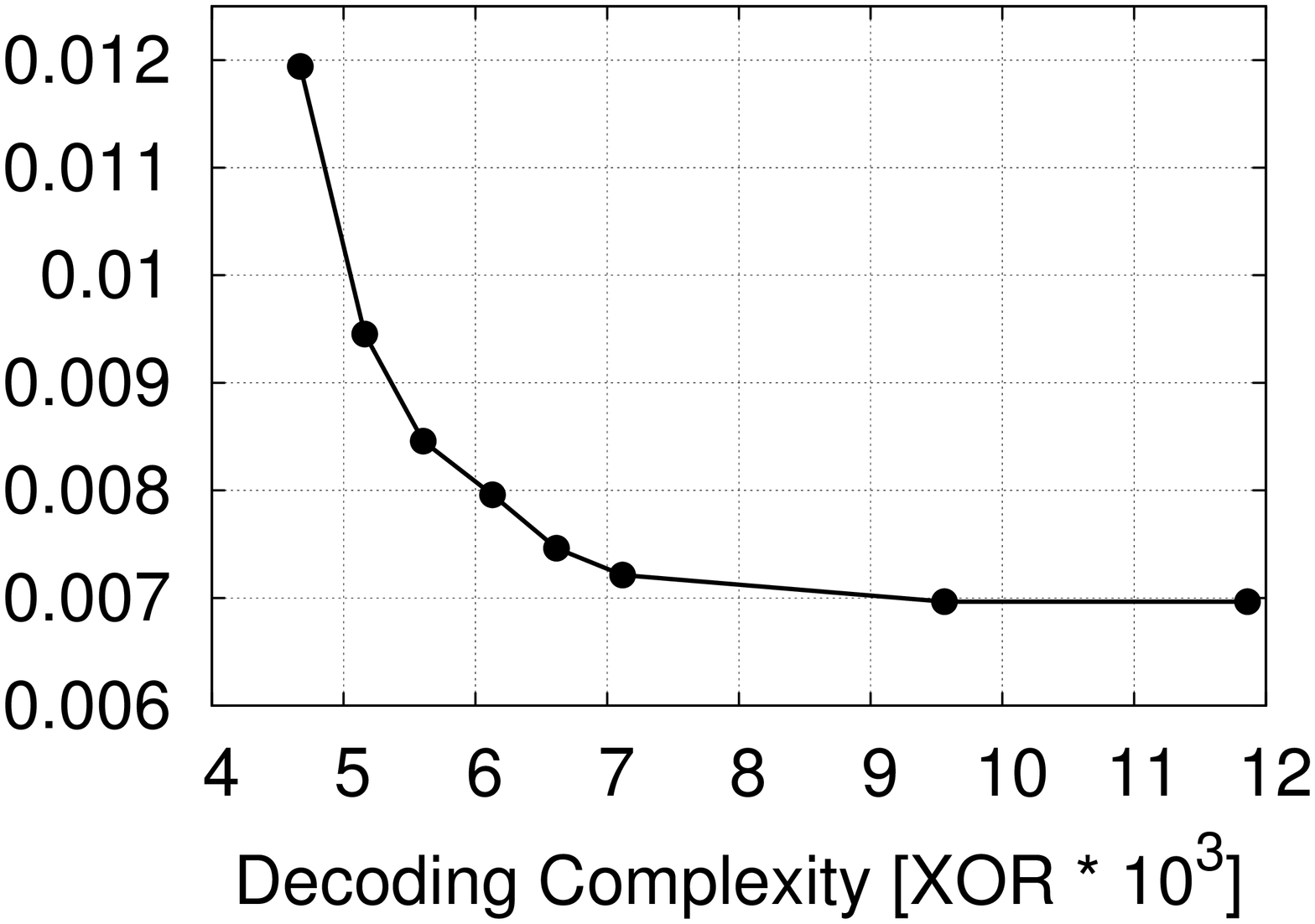}
	}
	}
	\caption{Source-receiver tradeoff between decoding complexity and encoding efficiency as a function of the encoding window size $\frac{W}{N} \in \{0.2,0.22,0.24,0.26,0.28,0.3,0.4,0.5\}$ for generations of N=100 (left) and N=200 (right) symbols.}
	\label{fig:complexity_overhead_source_receiver}
  \end{center}
  %\vspace{-10mm}
\end{figure}

In this subsection we measure the trade-off between decoding complexity and encoding efficiency of BC in a simple scenario where the source transmits encoded packets directly to the receiver, i.e. there are no recombinations in the network. 
%In this scenario, Band Codes behave like common rateless codes. 
The decoding complexity is measured as the number of XOR operations required to recover the generation as modeled in Section~\ref{sec:complexity_model}.
The encoding efficiency is measured in terms of \emph{encoding overhead}, which is defined as $\epsilon=\frac{N'}{N}-1$ and accounts for the extra bandwidth required to recover the video due to the linear dependencies between packets.
Figure~\ref{fig:complexity_overhead_source_receiver} shows the complexity-efficiency trade-off as a function of the ratio between the size of the encoding window $W$ and the size of the generation $N$.
The figure shows two curves, one for generations of $N = 100$ symbols (left) and the other for generations of $N=200$ symbols  (right).
Each curve is composed by 8 points that corresponds to $\frac{W}{N} \in \{0.2,0.22,0.24,0.26,0.28,0.3,0.4,0.5\}$ from top-left to bottom-right (our experiments showed that for $\frac{W}{N} > 0.5$ the encoding overhead does not decrease any further).
The figure shows that an encoding overhead of about 1.5\% and below 1\% respectively are possible, values comparable to other families of rateless codes such as LT or Raptor codes.
%However, for $W=N/2$ BC achieve the same efficiency as for $W=N$ but at half of the decoding complexity.

\subsection{Performance of BC for NC}

\begin{comment}
\begin{figure}[h]
  \begin{center}
	  \includegraphics[width=0.8\columnwidth]{figures/testbed.eps}
	\caption{The experimental testbed used to evaluate our P2P protocol for video streaming based on BC. The picture shows the source node node (left), the workstations that host the peer nodes (center) and the test mobile phone (right).}
	\label{fig:testbed}
  \end{center}
\end{figure}
\end{comment}

In this subsection we experiment in a more complex scenario that involves multiple receivers that recombine the received packets.
The experiments are performed on a controlled conditions testbed composed by several workstations connected via Ethernet links.% as shown in Figure~\ref{fig:testbed}.
One  workstation hosts the source node, while the remaining workstations host a total of 100 peer nodes.
Furthermore, an LG-P500 mobile phone powered by a 600 MHz ARM processor is connected to the rest of the testbed via a 802.11g WiFi link.
The nodes of the testbed form an overlay that implements our P2P protocol for live video streaming.
%We simulate a flash crowd scenario where the nodes join the overlay all at the same moment, each node is connected to a randomly selected subset of the peers in the overlay and the nodes in the overlay form a randomly connected graph with cycles.
We simulate a flash crowd scenario where the nodes join the overlay all at the same moment and form a fully connected overlay with cycles.
The source node streams a 10 minutes H.264/AVC video sequence that is subdivided in generations of $C_t = 1$ second each, and each generation is further subdivided in symbols of 1250 bytes.
Each generation encompasses one Group of Pictures (GOP), which is the minimum self-decodable unit of video in modern video coding standards such as H.264/AVC.
Enclosing one (or multiple) GOPs within the same generation guarantees that the video player is able to decode any recovered generation of the video independently from whether the adjacent generations were successfully recovered or not.
The buffering time is set to 5 seconds, i.e. the nodes start to play back the first buffered generation of video 5 seconds after they have entered the overlay.
The video sequence can be encoded at two different bitrates, namely 1 and 2 Mbit/s, which yields generations of $N=100$ and $N=200$ symbols respectively.
The output bandwidth of the source node is constrained so that, on average, each peer receives about 10\% of the packets from the source and about 90\% of the packets from the other peers.
%In our experiments we consider different sizes of the encoding window size $W$, which is the parameter that controls the tradeoff between decoding complexity and encoding efficiency of BC.
%In particular, we consider values of $$\frac{W}{N}$ \in [0.2, 0.4, 0.6, 0.8, 1.0]$, from the least complex and least efficient to the most complex and most efficient.
%The peer nodes are configured to request all the same encoding window size $W$ during the handshake procedure, so that we can study an homogeneous network of peers.
%The experiment is repeated multiple times, and at each iteration the nodes are configured to request a different $W$ to their peers during the handshake process.
%For each experimental configuration, we measure the resulting computational complexity and processor load in the mobile node.

\begin{figure}[ht!]
	\begin{center}
	\subfigure{
		\rotatebox{270}{\includegraphics[width=0.40\columnwidth]{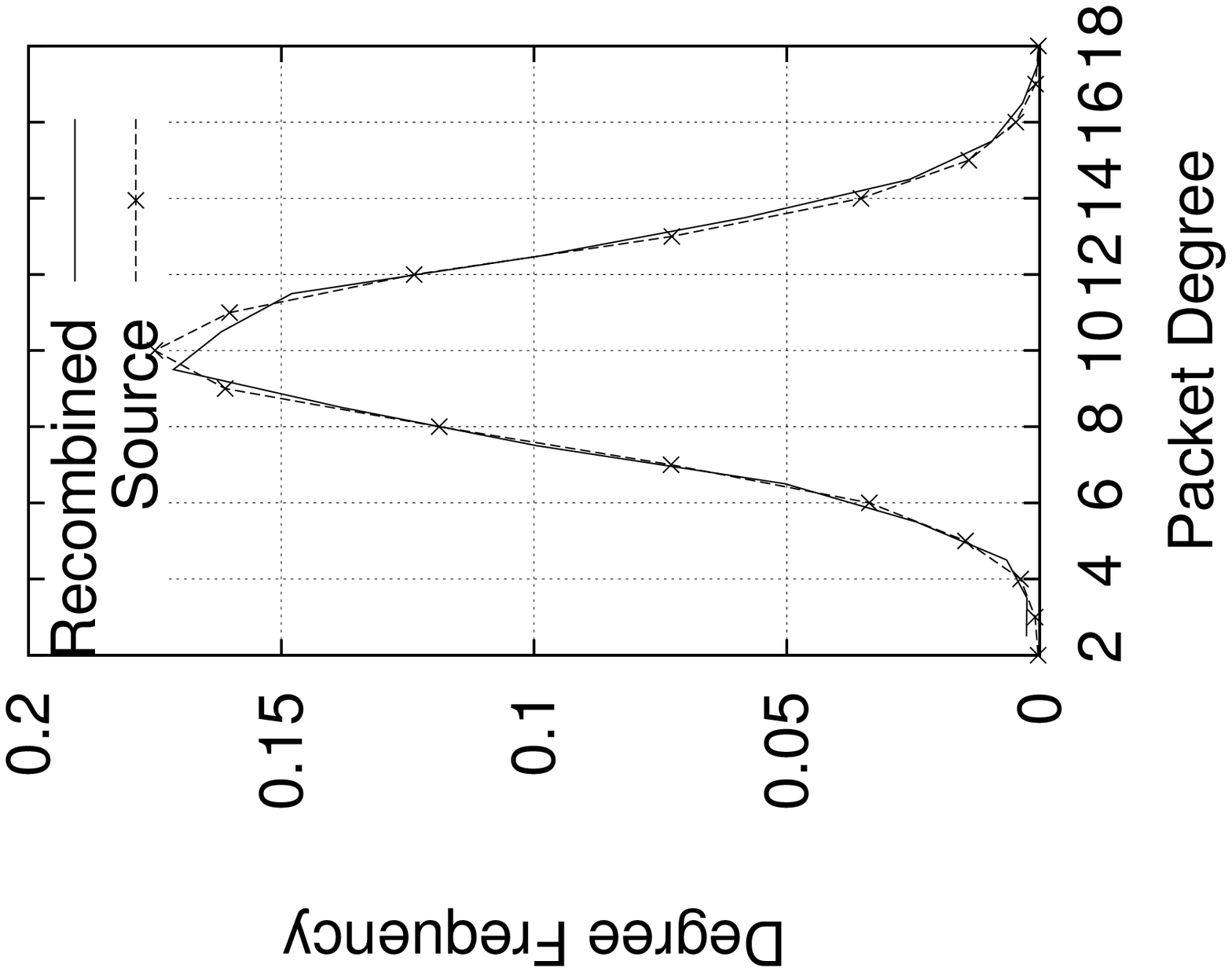}}
	}
	\subfigure{
		\rotatebox{270}{\includegraphics[width=0.40\columnwidth]{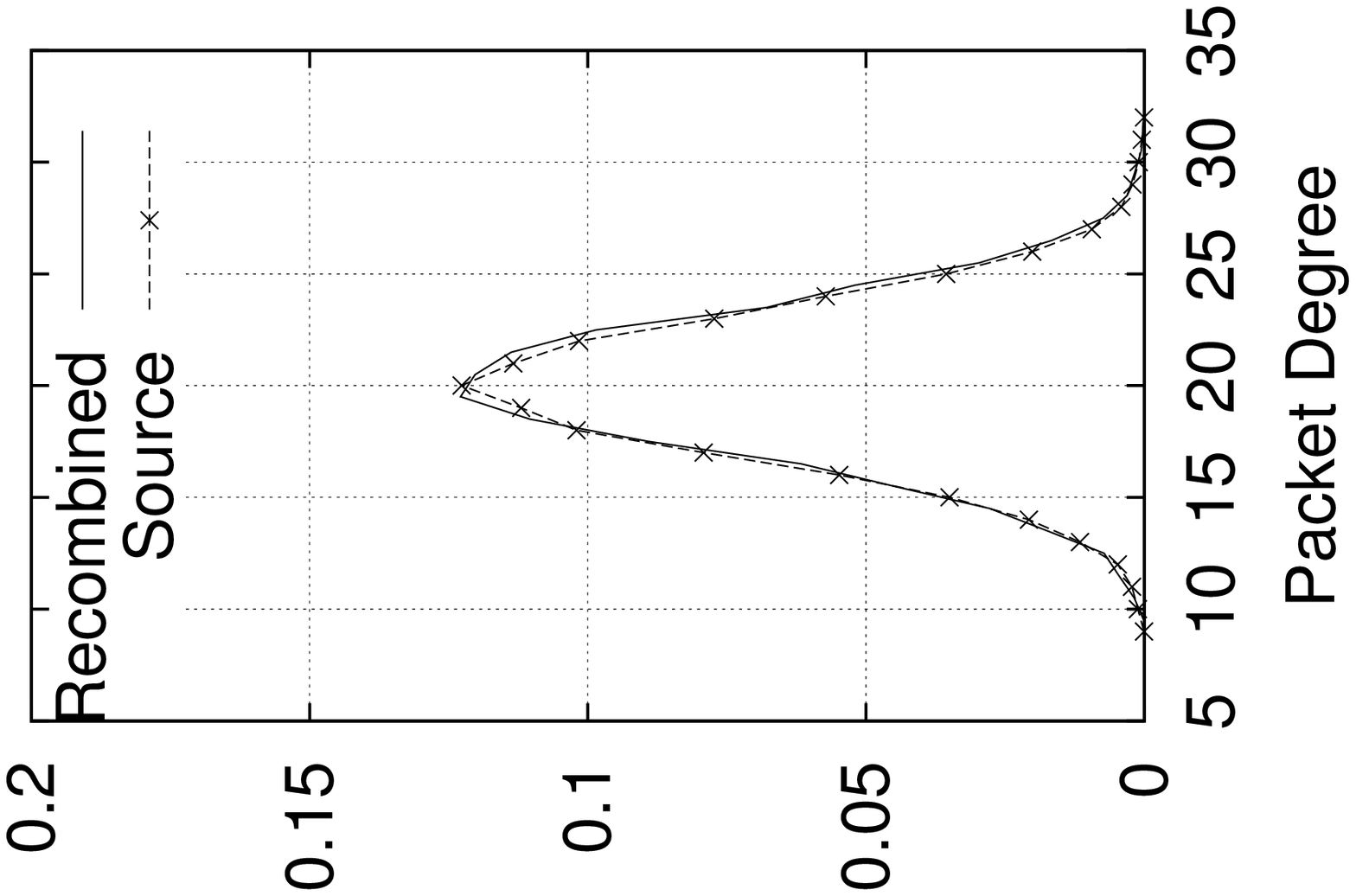}}
	}
	\subfigure{
		\rotatebox{270}{\includegraphics[width=0.40\columnwidth]{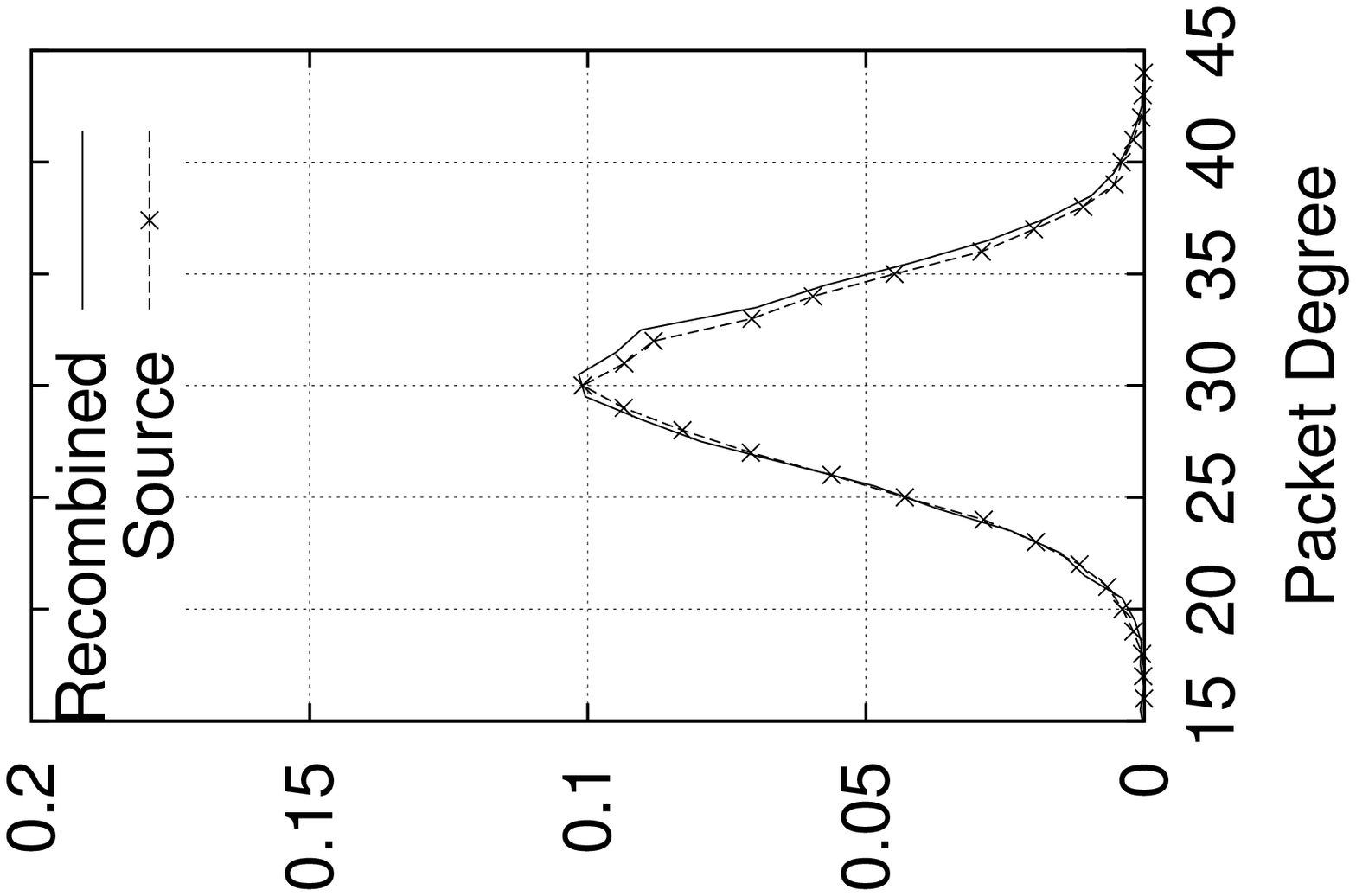}}
	}
	\end{center}
	\caption{Packet degree distribution for an encoding window size $\frac{W}{N}$ equal to 0.2 (left), 0.4 (center), 0.6 (right) for a generation of N=100 symbols. The recombinations at the network nodes do not alter the distribution imposed at the source node.}
	\label{fig:degree_distribution}
\end{figure}

Figure~\ref{fig:degree_distribution} shows the degree distribution of the packets received by the network nodes after the recombinations for the 1 Mbit/s video sequence ($N = 100$) and for different encoding windows of size $\frac{W}{N} \in \{0.2, 0.4, 0.6\}$.
The figure also shows the degree distribution imposed by the source node, i.e. the Binomial Distribution $\mathcal{B}(W,\frac{1}{2})$.
The packet degree distribution at the network nodes closely follows the degree distribution imposed at the source, showing that BC preserves the degree distribution imposed at the source despite the recombinations.

\begin{comment}
\begin{figure}[h]
  \begin{center}
	\rotatebox{270} {
	  \includegraphics[width=0.5\columnwidth]{figures/degree_distribution.eps}
	}
	\caption{Distribution of the degree of the packets in the network for $N=100$ for Band Codes.}
	\label{fig:degree_distribution}
  \end{center}
  \vspace{-10mm}
\end{figure}
\end{comment}

\begin{figure}[ht!]
	\begin{center}
	\subfigure{
		\rotatebox{270}{\includegraphics[width=0.305\columnwidth]{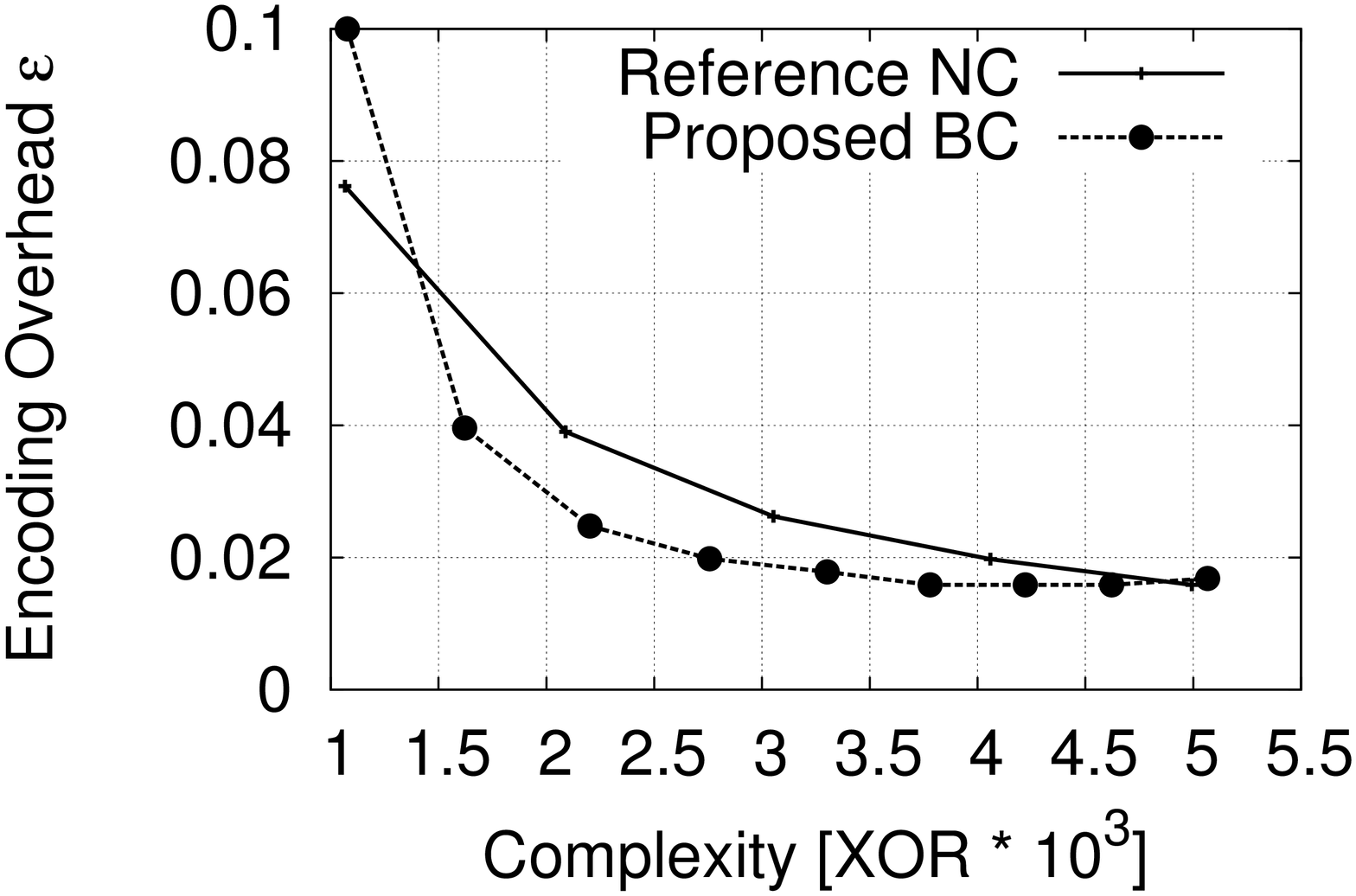}}
	}
	\subfigure{
		\rotatebox{270}{\includegraphics[width=0.305\columnwidth]{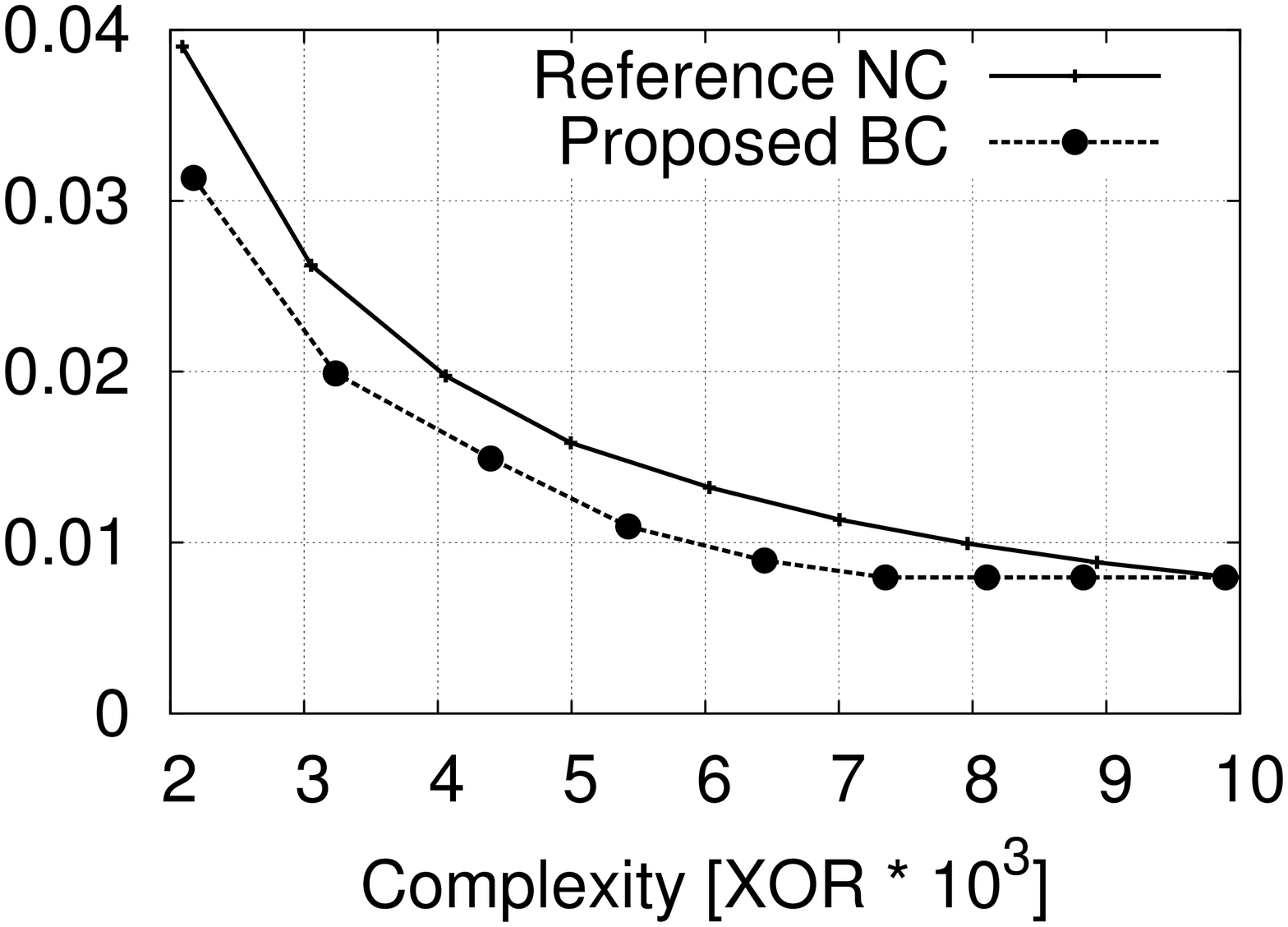}}
	}
	\end{center}
	\caption{Tradeoff between decoding complexity and encoding efficiency as a function of the encoding window size $\frac{W}{N} \in \{0.2,0.3,0.4,0.5, 0.5, 0.7, 0.8, 0.9, 1.0\}$ for generations of N=100 (left) and N=200 (right) symbols.}
	\label{fig:complexity_overhead}
\end{figure}

%\subsection {Encoding Efficiency of Band Codes}
Then, we move to study the tradeoff between decoding complexity and encoding efficiency of  BC.
Figure~\ref{fig:complexity_overhead} shows two sub-figures, one for generations of $N$ = 100 symbols (left) and one for generations of $N$=200 symbols (right).
The \emph{Proposed-BC} curve is composed by 9 points corresponding to encoding windows of size $\frac{W}{N}\in \{0.2, 0.3, 0.4, 0.5, 0.6, 0.7, 0.8, 0.9, 1.0\}$ from top-left to bottom-right.
The \emph{Reference NC} curve shows the complexity-overhead tradeoff of a standard NC scheme as described in Section~\ref{sec:rnc_over_gf2}.
In this latter reference scheme, the generation size $N$ is gradually reduced from 100 to 20 symbols in the left figure and from 200 to 40 symbols in the right figure to control the decoding complexity.
For each point on the curve, we show the number of XOR operations required to recover 1 Mbit of encoded payload and the corresponding overhead.
As expected, when $\frac{W}{N}=1$ (bottom-right point in the figures) BC perform as a standard NC scheme.
Otherwise, BC enables better complexity-overhead tradeoff than standard NC, as each Proposed BC curve is below the corresponding Reference NC curve in almost any situation.
That is, for a given decoding complexity BC enable lower overhead than the reference scheme (and the other way around).
For $N$=100 and $\frac{W}{N}=1$, the encoding overhead is about 1.6\% and the decoding complexity is  about 5100 XOR.
The comparison with Figure~\ref{fig:complexity_overhead_source_receiver} shows that the loss in encoding efficiency of BC due to the recombinations at the nodes is less than 1\% when $\frac{W}{N}=1$.
When $\frac{W}{N}$=0.5, the decoding complexity of BC drops by a factor of two with an encoding overhead penalty below 0.5\%. 
For a generation of $N$=200 symbols, similar results hold: the decoding complexity drops by a factor of two with an overhead penalty of just 0.5\% when $\frac{W}{N}=0.5$ with respect to the reference case $\frac{W}{N}=1$.
In both cases, an encoding overhead of about 5\% is possible with a reduction in the decoding complexity of nearly four times.
\begin{comment}
\\
The figure also shows the computational complexity of the recombination process $C_R$, which is the average number of XOR operations performed by a node to recombine the rows of $G$ per each generation.
While modeling the recombination complexity proved to be quite complex, we provide an intuitive explanation and an experimental evidence of how BC reduces it.
Let $\frac{|R|}{2}$ be the average number of equations recombined at a given transmission opportunity in Algorithm~\ref{alg:recombination}.
The average number of XOR operations required to recombine them is $\frac{|R|}{2} -1$, where $|R|$ directly depends on $W$: as the encoding window increases, more equations are in fact suitable for recombination.
Therefore, the number of XOR operations $C_R$ performed by a node to recombine the rows of $G$ for each generation increases with $W$.
The figure shows that $C_R$ decreases by a factor of two already when $\frac{W}{N}$=0.9 and by a factor of five when $\frac{W}{N}$=0.5 with little impact on the encoding efficiency.
Thus, as both the decoding and recombination process contribute to the actual computational complexity of NC, we conclude that in our experiments we were able to reduce the overall computational complexity of NC by a factor in excess of two without appreciable losses in encoding efficiency with respect to standard random NC.
\end{comment}

\begin{figure}[ht!]
	\begin{center}
	\subfigure{
		\rotatebox{270}{\includegraphics[width=0.33\columnwidth]{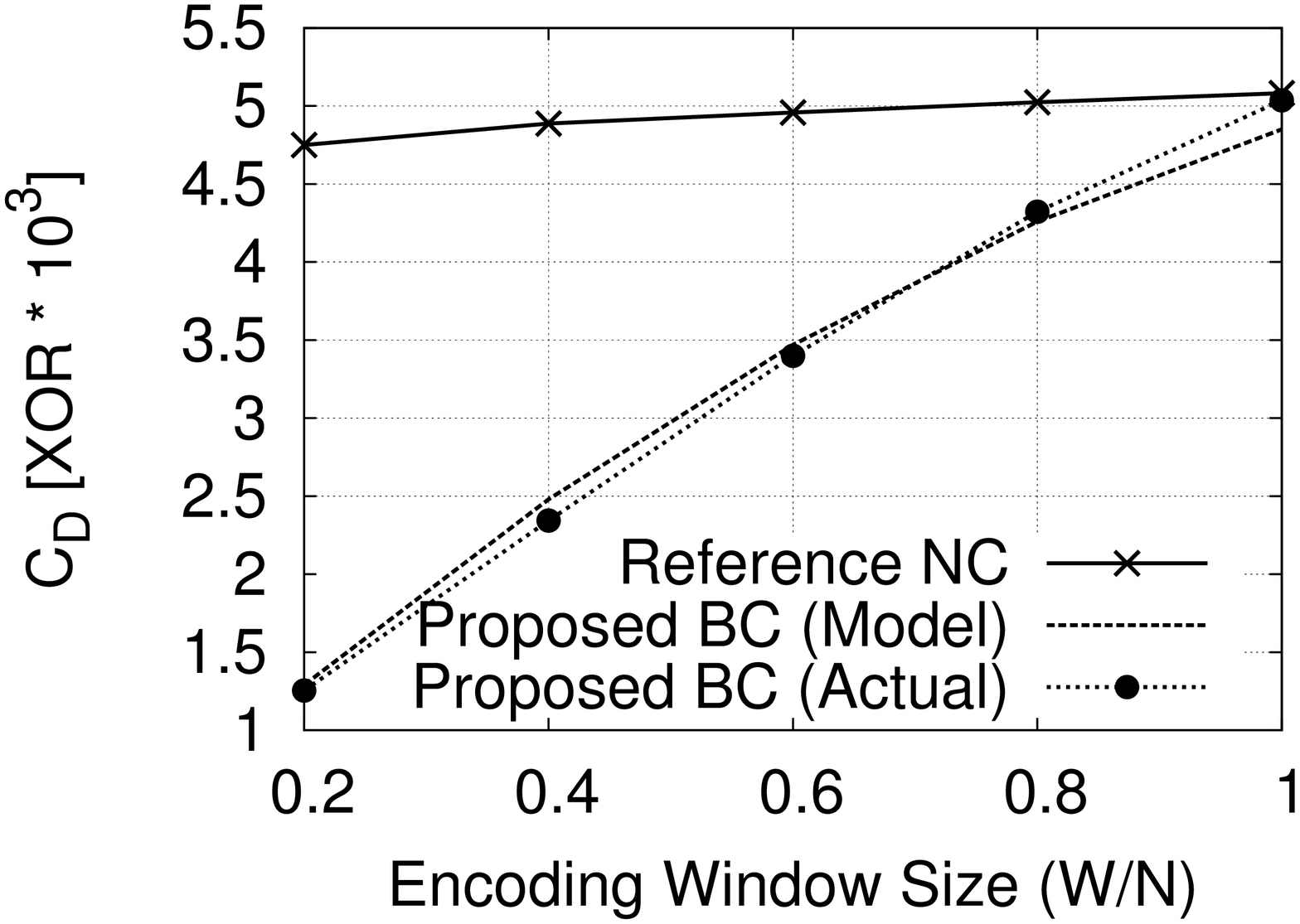}}
	}
	\subfigure{
		\rotatebox{270}{\includegraphics[width=0.33\columnwidth]{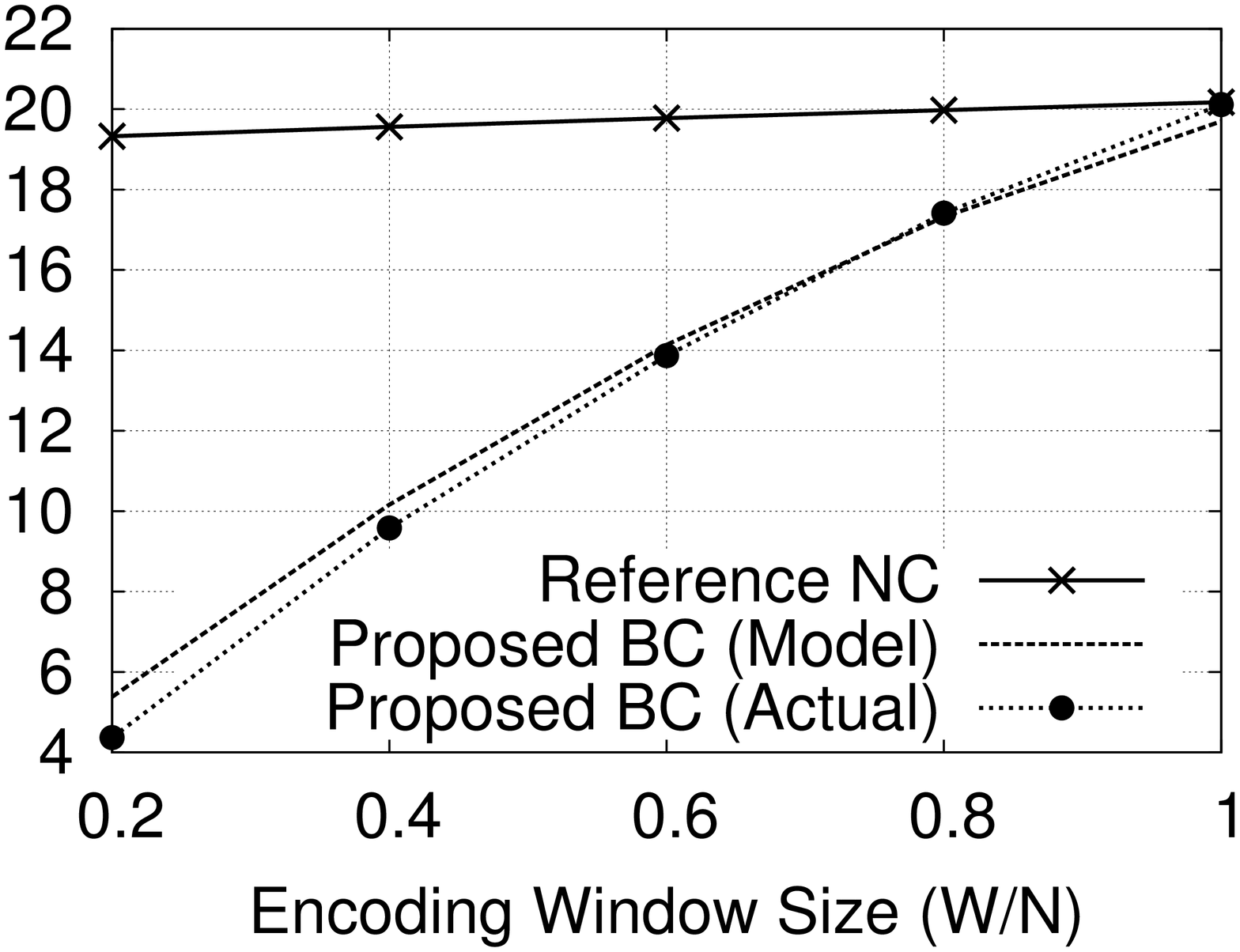}}
	}
	\end{center}
	\caption{Actual and predicted decoding complexity as a function of the encoding window size $\frac{W}{N} \in \{0.2,0.4,0.6,0.8,1.0\}$ for generations of N=100 (left) and N=200 (right) symbols.}
    \label{fig:decoding_cost}
\end{figure}

Figure~\ref{fig:decoding_cost} shows the actual and predicted decoding complexity of BC for different values of the encoding window size $W$.
As a reference, we consider a scheme where the network nodes transmit random linear combinations of the received packets as described in Section~\ref{sec:rnc_over_gf2}.
We see that the decoding complexity of BC grows linearly with $W$ as accurately predicted by the model in~(\ref{eqn:cost_decoding}).
On the contrary, the decoding complexity of the reference scheme remains close to the asymptotic value as it increases with the number of recombinations in the network.

%\ifDRAFTMODE
\begin{figure}[ht!]
	\begin{center}
	\subfigure{
		\rotatebox{270}{\includegraphics[width=0.33\columnwidth]{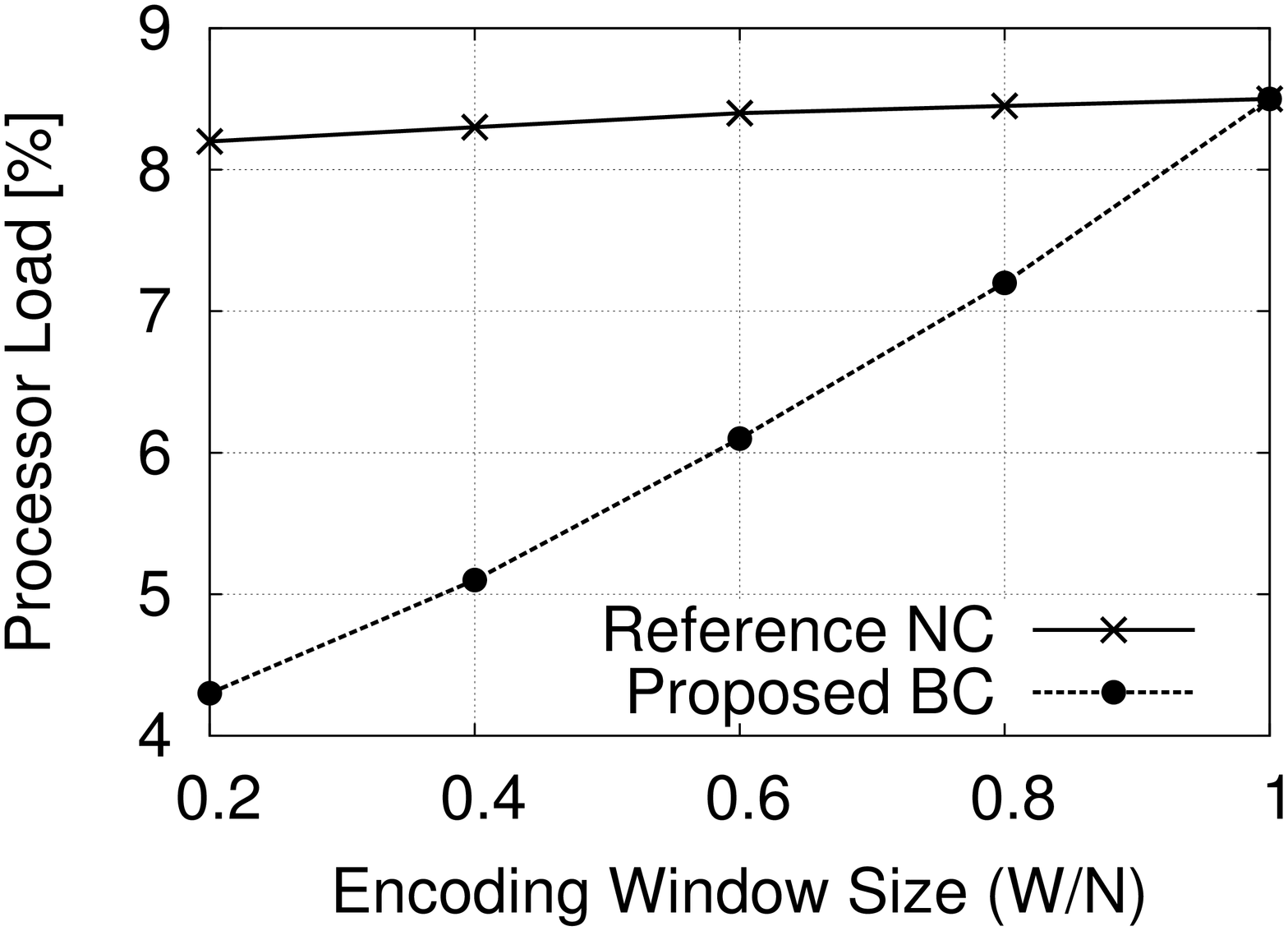}}
	}
	\subfigure{
		\rotatebox{270}{\includegraphics[width=0.33\columnwidth]{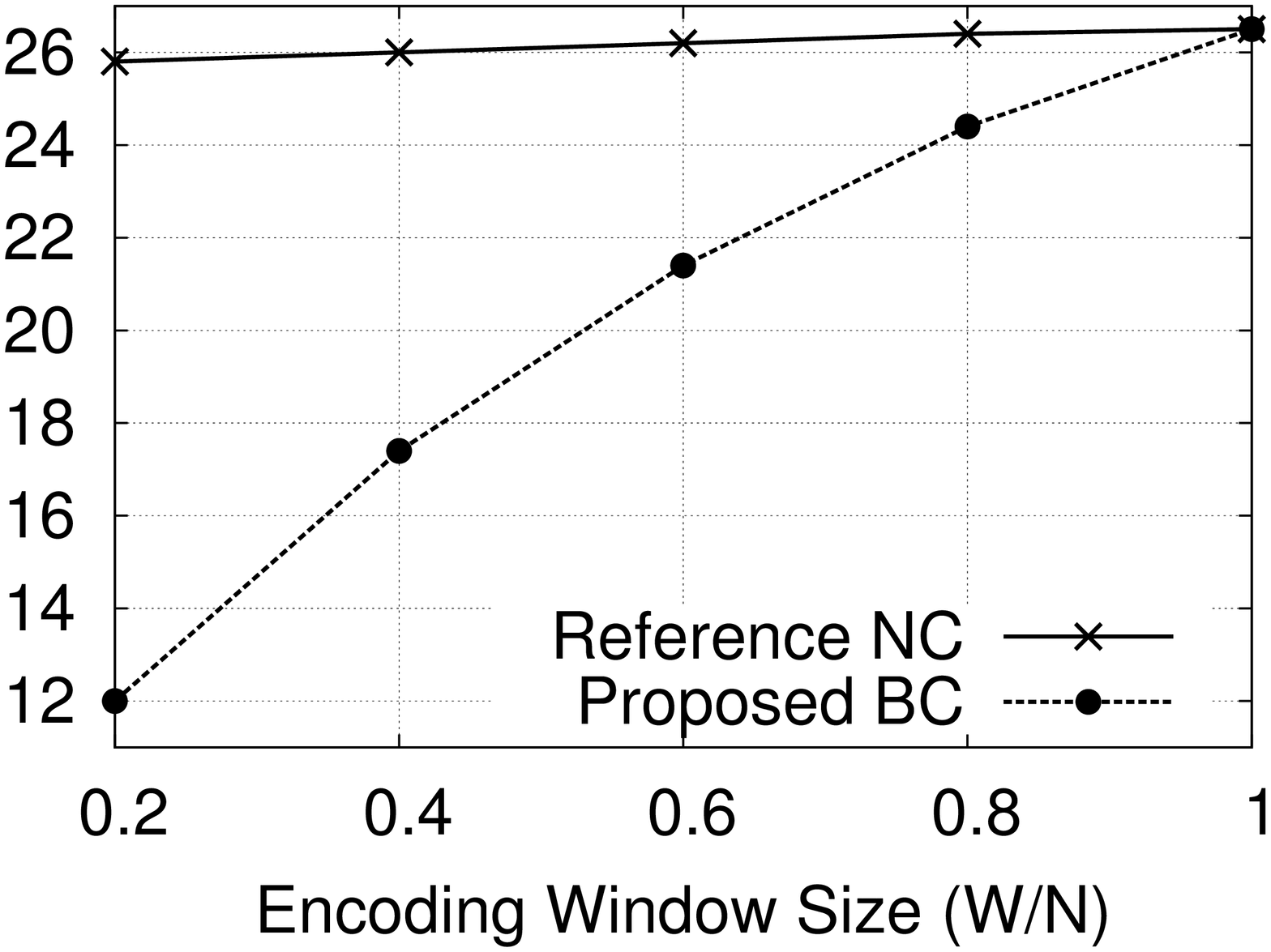}}
	}
	\end{center}
	\caption{Processor load on the mobile phone as a function of the encoding window size $\frac{W}{N} \in \{0.2,0.4,0.6,0.8,1.0\}$ for generations of N=100 (left) and N=200 (right) symbols.}
    \label{fig:android_cpu}
\end{figure}

\begin{comment}
\else
\begin{figure}[h]
  \begin{center}
    \rotatebox{270}{\includegraphics[width=0.5\columnwidth]{figures/decoding_cost_multiplot.eps}}
    \caption{Decoding complexity $C_D$ as a function of the encoding window size $\frac{W}{N}$. Video bandwidth is 1 Mbps (top) and 2 Mbps (bottom.) Lower figures are better.}
    \label{fig:decoding_cost}
  \end{center}
%\vspace{-20pt}
\end{figure}
\fi
\end{comment}
Figure~\ref{fig:android_cpu} shows the corresponding processor load measured at the mobile node. % corresponding to the decoding complexity shown in Figure~\ref{fig:decoding_cost}.
A comparison with Figure~\ref{fig:decoding_cost} shows the correlation between the decoding complexity and the actual processor load.
In particular, a reduction in the decoding complexity results in a quasi proportional reduction of the processor load.

\begin{figure}[h]
  \begin{center}
	\rotatebox{270}{\includegraphics[width=0.5\columnwidth]{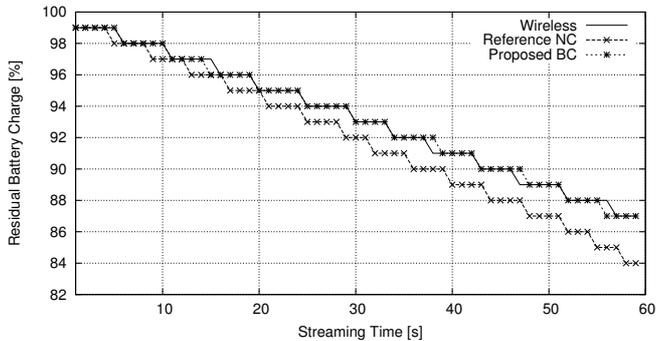}}
	\caption{Residual battery charge over time. BC reduces the energy consumption of the network nodes and extends battery life.}
	\label{fig:battery_charge}
  \end{center}
%    \vspace{-5mm}
\end{figure}

Next, we measure the actual energy consumption of the mobile phone as a function of three different NC strategies.
The first scheme $Proposed~BC$ corresponds to the BC scheme proposed in this paper, where the packets in the network are encoded with $\frac{W}{N}$=0.5. % and the nodes recombine a maximum of $N_r = 4$ equations at each transmission.
The second scheme $Reference~NC$ represents the case where the network nodes transmit random linear combinations of the received packets as described in Section~\ref{sec:rnc_over_gf2}.
With respect to the first scheme, the packet decoding and recombination process is more complex and thus is expected to consume more energy.
Finally, we consider a third scheme, $Wireless$, where we send UDP traffic to the mobile node at the rate of 2 Mbit/s without NC.
This scheme has the purpose of assessing and setting off the power consumption of the mobile phone due to the access to the wireless channel.
%This latter scheme enables us to determine  the energy consuption of the sole compare the interface leaving out the contribution of the coding operations.
We stream six times in loop the same 10 minutes video sequence encoded at 2 Mbps ($N=200$) used in the previous experiments, for a total streaming time of 60 minutes.
The residual level of battery charge is logged at intervals of 60 seconds using the monitoring interface provided by the Android operating system.
The experiment is executed once for each of the three considered schemes, and the battery of the mobile node is fully recharged before each experiment.
Figure~\ref{fig:battery_charge} shows the residual level of battery charge as a function of the streaming time.
The reference scheme $Wireless$ results in a battery discharge equal to 13\% due to the access to the wireless channel.
The scheme $Reference~NC$ results in a battery discharge level equal to 16\%, that is the energy consumption increase with respect to the $Wireless$ scheme is equal to 23\% and is due to the coding operations at the network nodes.
Finally, the scheme $Proposed~BC$ results in a discharge level equal to 14\%, that is only 7\% greater that the scheme that considers just the access to the communication channel.
The experiment shows that BC reduces the energy consumption due to the coding operations from 23\% to 7\% with respect to the sole access to the wireless channel, with a reduction of energy consumption larger than a factor of 3 with respect to a conventional random linear NC scheme. 

\subsection {Video Quality Assessment over PlanetLab}

We experiment with our P2P protocol on the PlanetLab testbed~\cite{chun2003planetlab} to measure the quality of the video received at the network nodes.
PlanetLab is a testbed composed of hundreds of Internet nodes that makes it possible to experiment in a network scenario where packet erasures, delays and out of order delivery impair the quality of the streaming session.
%We previously~\cite{fiandrotti2011complexity} used PlanetLab to compare a preliminary version of the P2P protocol presented in this paper with the NextShare protocol~\cite{nextshare}.
%Our previous experiments showed that our NC-based P2P protocol achieved slightly better performance with respect to NextShare but with reduced buffering time, hence it can be considered as a reference of the state-of-the-art of the binary NC protocols. 
In~\cite{fiandrotti2011complexity} PlanetLab is used to compare a preliminary version of the P2P protocol presented in this paper with the NextShare protocol~\cite{nextshare}.
The experiments showed that the NC-based P2P protocol proposed in~\cite{fiandrotti2011complexity} achieved slightly better performance with respect to NextShare but with reduced buffering time, hence it can be considered as a reference of the state-of-the-art of the binary NC protocols. 

\begin{figure}[h]
  \begin{center}
	\rotatebox{270} {
	  \includegraphics[width=0.5\columnwidth]{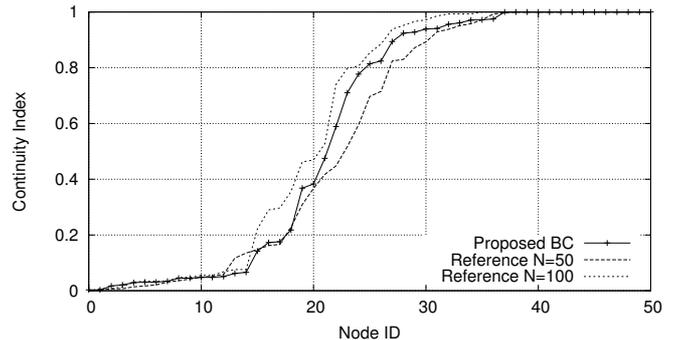}
	}
	\caption{Continuity index measured at the PlanetLab nodes for different complexity control strategies (the 50 nodes with lower CI are shown).}
	\label{fig:planetlab_ci}
  \end{center}
  \vspace{-5mm}
\end{figure}

%NextShare~\cite{nextshare} represents the state-of-the-art in P2P video distribution non designed around the NC paradigm.
%The comparison was meant to assess the pros and the cons of a P2P protocol designed around NC over a traditional protocol non designed around NC.
%The experiments showed that the simplified version of the protocol performed as well as the much more refined NextShare.
%In this paper we experiment how the energy-aware features of our P2P protocol affect the quality of the video received at the nodes.
The video quality is measured in terms of Continuity Index (CI), which represents the percentage of generations that could be correctly recovered before the playout deadline.
The experiments are performed using 300 randomly selected PlanetLab nodes plus one node located at the Politecnico di Torino that acts as source node.
%The experiments are performed at night, when the volume of time-varying, interfering, traffic is lower.
We stream a 10 minutes H.264/AVC test video sequence encoded at 500 kbps and the temporal duration of each GOPs is 2 seconds, i.e. each GOP is 1 Mbit. 
We experiment with three different parameters, namely the complexity control strategy, the source node bandwidth $B_s$ and the initial buffering time $t_b$.
%We experiment with three different parameters of our P2P protocol, namely the size $W$ of the encoding window, the bandwidth of the source node $B_s$ and the initial buffering time $t_b$.
\\
Figure~\ref{fig:planetlab_ci} shows the CI measured at the PlanetLab nodes.
The bandwidth of the source node and the peer nodes are respectively set $B_s = 5$ Mbit/s and $B_p$ = 750 Kbit/s, while the buffering time is $t_b$ = 10 s.
The figure contains three curves that show the CI delivered by three different decoding complexity control schemes.
The \emph{Reference - N 100} curve is for a random NC scheme where each generation is $N$=100 symbols and contains a whole GOP and the corresponding decoding complexity is about 5000 XOR per GOP.
The \emph{Reference - N 50} curve is for a random NC scheme where each generation is $N$=50 symbols and each GOP is split in two generations.
The first generation that compose a GOP can be independently decoded while the second can be decoded only if the first was recovered as well.
The decoding complexity of this scheme is 1225 XOR per generation, i.e. 2450 XOR per GOP.
The \emph{Proposed BC} curve is for a NC scheme based on BC where each generation is N=100 symbols and contains a whole GOP.
The encoding window size is equal to W=37 symbols, which entails a decoding complexity of 2400 XOR per GOP, comparable to the complexity of the Reference - N 50 scheme.
As expected, the \emph{Reference - N 100} curve shows the best video quality (average CI equal to 0.930) in reason of the lowest overhead.
The \emph{Reference - N 50} scheme achieves a CI equal to 0.921 in reason of the slightly higher code overhead and in reason of the fact that some generations could not be decoded because the first generation was lost.
Finally, the \emph{Proposed BC} scheme achieves a CI (0.925) that is lower than the “Reference - N 100” scheme but is slightly higher than the \emph{Reference - N 50} scheme.
That is due to the fact that the encoding overhead is slightly better for generations of N=100 symbols and each generation can be independently decoded.

\begin{figure}[h]
	\begin{center}
	\subfigure{
		\rotatebox{270}{\includegraphics[width=0.33\columnwidth]{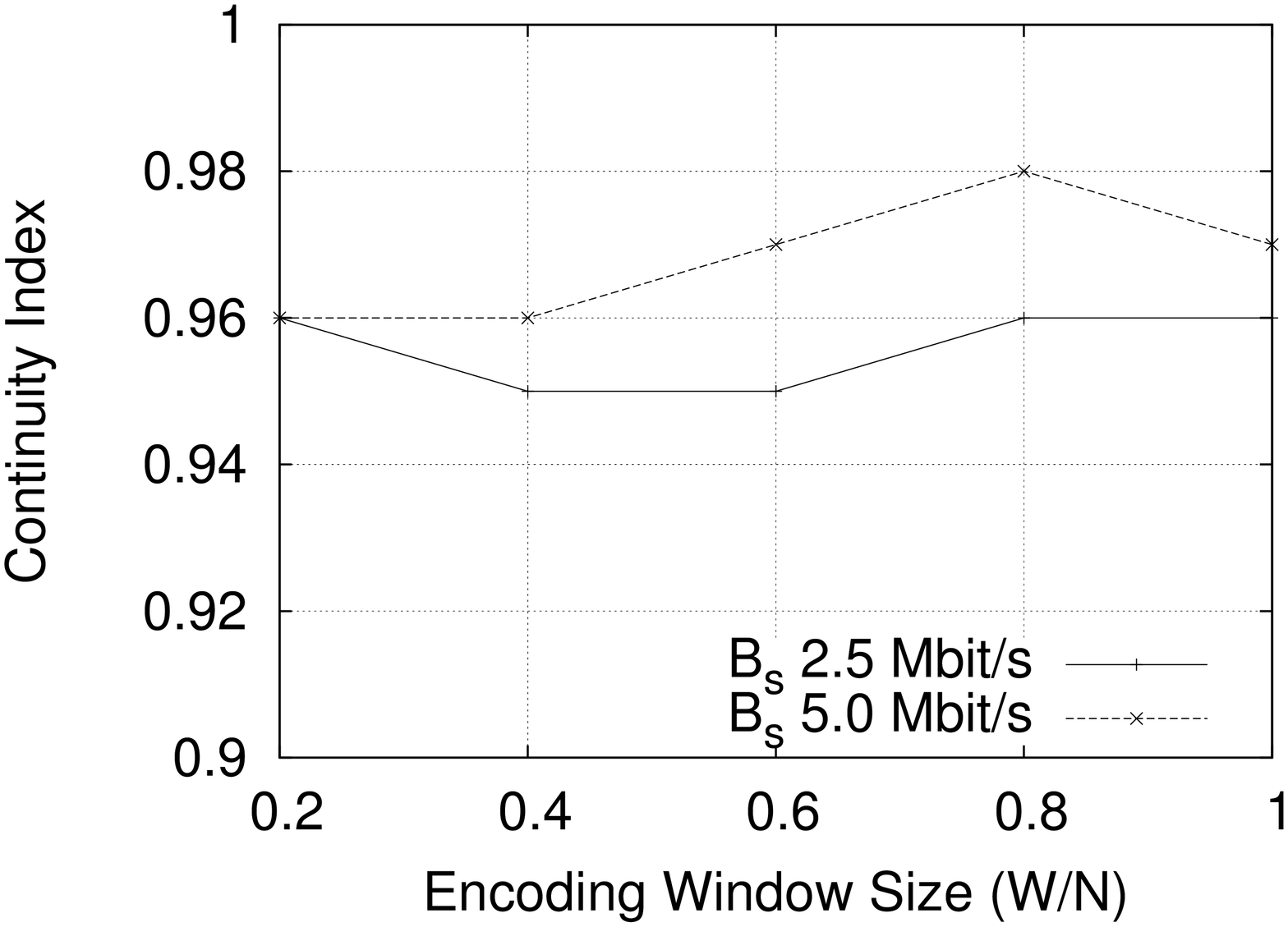}}
%		\label{fig:planetlab_bs}
	}
	\subfigure{
		\rotatebox{270}{\includegraphics[width=0.33\columnwidth]{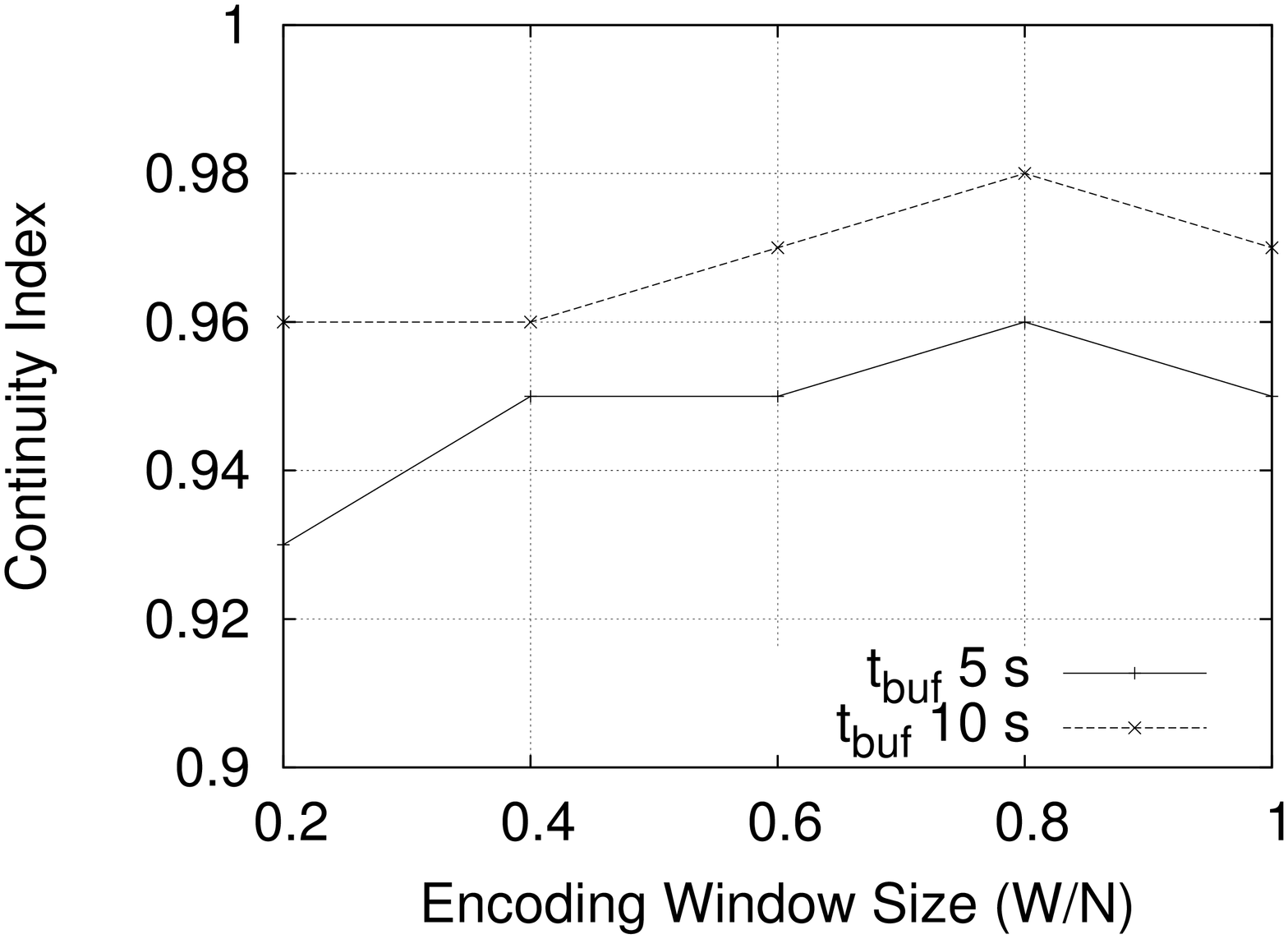}}
%		\label{fig:planetlab_tbuf}
	}
	\end{center}
	\caption{Average CI measured at the PlanetLab nodes as a function of the encoding window size ($\frac{W}{N}$). We reduced the source bandwidth $B_s$ from 5 to 2.5 Mbit/s (left) and the initial buffering time $t_b$ from 10 to 5 s (right).}
    \label{fig:planetlab_bs_tbuf}
\end{figure}

We repeat the previous experiment reducing first the bandwidth $B_s$ of the source node from 5 to 2.5 Mbit/s for $t_b$ = 10 s and then reducing the buffering time $t_b$ from 10 to 5 seconds for $B_s$ = 5 Mbit/s.
Such experiments aim at stressing the P2P protocol by reducing the available output bandwidth and assessing the effect on the quality of the streaming session.
Figure~\ref{fig:planetlab_bs_tbuf} shows the results of the experiments in terms of average CI measured at the network nodes.
As $B_s$ decreases, the upload bandwidth available in the network decreases, so small reduction in the continuity index is observed (between 1 and 2 \%).
The analysis of the logs showed that the nodes that suffer the most from the reduced source bandwidth are those more affected by packet losses, which correspond to the points located in the leftmost part of Figure~\ref{fig:planetlab_ci}.
However, the experiment suggests that the source node is able to support up to 150 peers feeding each node with less than 3.3\% of the bandwidth required to decode a generation keeping into account also the encoding overhead.
A reduction in $t_b$ results in reduced initial delay for the user but also in reduced likelihood to put up with the packet losses on the network.
The CI reduction ranges between 1\% and 3\%, however our architecture is still able to achieve an average CI greater than 90\% (86\% of the nodes boast a CI higher than 90\%) despite the reduced buffering time and the insufficient bandwidth of some of the PlanetLab nodes.

\begin{comment}
\begin{figure}[h]
  \begin{center}
	\rotatebox{270} {
	  \includegraphics[width=0.4\columnwidth]{figures/planetlab_ci_bs.eps}
	}
	\caption{Average CI measured at the PlanetLab nodes as a function of the encoding window size $\frac{W}{N}$ and bandwidth of the source node $B_s$. Higher figures are better.}
	\label{fig:planetlab_bs}
  \end{center}
  \vspace{-5mm}
\end{figure}
\end{comment}

\begin{comment}
\begin{figure}[h]
  \begin{center}
	\rotatebox{270} {
	  \includegraphics[width=0.4\columnwidth]{figures/ci_tbuf_planetlab.eps}
	}
	\caption{Average CI measured at the PlanetLab nodes as a function of the encoding window size $\frac{W}{N}$ and initial buffering time $t_b$. Higher figures are better.}
	\label{fig:planetlab_tbuf}
  \end{center}
  \vspace{-5mm}
\end{figure}
\noindent
\end{comment}

\section{Conclusions and Future Work}
\label{sec:conclusions}

In this paper we presented Band Codes (BC), a family of codes that preserve the packet degree distribution enabling controlled-complexity NC independently from the network topology.
Our experiments show that BC reduce the decoding complexity by a factor of two with almost no losses in encoding efficiency with respect to random NC.
Furthermore, a reduction of up to four times is achieved while maintaining the encoding overhead below 5\%.
Experiments with a mobile phone showed that the reduced computational complexity reduces its energy consumption extending the operational lifetime.
Streaming experiments show that our P2P protocol designed around BC is capable to deliver high quality video on the global scale testbed, showing the benefits of BC in a realistic setting.
Although in this work we have focused on P2P video streaming,  BC are well suited in any scenario where energy consumption is a critical issue, such as sensor networks.
Finally, while we have considered NC over $GF(2)$ thanks to its low complexity, the main concepts behind BC can be extended to Galois fields of larger size.

%\section*{Acknowledgments}

%This publication is based on work performed in the framework of the Project COAST-ICT-248036, which is partially funded by the European Union.

\appendix[Proof of Proposition \ref{prop:recombination_issue}]
\label{sec:appendix}
We indicate with $\Omega^j$ the packet degree distribution in the network after $j$ recombinations, i.e. the probability that a randomly selected packet in the network has degree $i$ after $j$ recombinations is $\Omega^j_i$.
The source node encodes packets of degree $d$ with probability $\Omega_d^0$.
Let $P^1$ and $P^2$ be two packets in the network with degree $d^1$ and $d^2$.
We define as $s_N(d^1, d^2, d^r)$ the probability that the recombination of $P^1$ and $P^2$ produces a packet $P^r$ with degree $d^r = d^1 + d^2 - 2\chi$, where $\chi$ is the random variables that counts the number of times that $g_i^1 = g_i^2 = 1$ for $i \in [0, N-1]$.

\ifDRAFTMODE
	\begin{equation*}
	s_N (d^1, d^2, d^r) = s_N (d^1, d^2, d^1 + d^2 - 2\chi) = \mathbb{P}(2\chi = d^1 + d^2 - d^r) = \mathbb{P}(\chi = \frac{ d^1 + d^2 - d^r }{2}).
	\end{equation*}
\else
	\[\begin{array}{r c l}
	s_N (d^1, d^2, d^r)& = & s_N (d^1, d^2, d^1 + d^2 - 2\chi)\\
	& = & \mathbb{P}(2\chi = d^1 + d^2 - d^r)\\
	& = & \mathbb{P}(\chi = \frac{ d^1 + d^2 - d^r }{2})\\
	\end{array}
	\]
\fi
\noindent
As $\chi$ follows the Hypergeometric Distribution $\mathcal{H}(N, d^1, d^2)$, we rewrite the above equation as

\[\begin{array}{r c l}
s_N (d^1, d^2, d^r)& = & \frac{\binom{d^1}{\frac{ d^1 + d^2 - d^r }{2}}\binom{N-d^1}{d^2-\frac{ d^1 + d^2 - d^r }{2}}}{\binom{N}{d^2}}.
\end{array}
\]
%\vspace{-5mm}
For the law of total probability, we have that
\\
\begin{equation}
\label{eqn:omega_recursive}
\Omega^j_i = \sum_{d^1=0}^{N} \sum_{d^2=0}^{N} s_N(d^1,d^2,i) \Omega_{d^1}^{j-1} \Omega_{d^2}^{j-1}.
\end{equation}
\noindent
%We define as the \emph{asymptotic distribution} $\Omega^\infty$ the distribution of the degree of the packets in the network after a number of recombinations that tends to infinite.
We indicate as $\Omega^\infty$ the distribution of the degree of the packets in the network after a number of recombinations that tends to infinite.
If $\Omega^0$ is not degenerate, we have from Equation~\ref{eqn:omega_recursive} that
\[\Omega^{ \infty }_i  =\frac{\binom{N}{i}}{2^{N}}.\]
\noindent
Therefore, the distribution of the degree of the packets in the network follows the Binomial Distribution $\mathcal{B}(N, \frac{1}{2})$ and the average degree of the packets in the network tends to $\frac{N}{2}$.

% or


\begin{comment}
\section*{Acknowledgment}

This publication is based on work performed in the framework of the Project COAST-ICT-248036, which is partially funded by the European Community.
\end{comment}

\ifCLASSOPTIONcaptionsoff
  \newpage
\fi



%\bibliographystyle{IEEEtran}

\bibliographystyle{IEEEtran}
\bibliography{main}

%\begin{biography}[{\includegraphics[width=1in,height=1.25in,clip,keepaspectratio]{mshell}}]{Michael Shell}

\begin{comment}

\begin{IEEEbiography}{Michael Shell}
Biography text here.
\end{IEEEbiography}

\begin{IEEEbiographynophoto}{John Doe}
Biography text here.
\end{IEEEbiographynophoto}

%\newpage

\begin{IEEEbiographynophoto}{Jane Doe}
Biography text here.
\end{IEEEbiographynophoto}
\end{comment}


\vfill


\end{document}